\documentclass[%
 aip,
jcp,%
 amsmath,amssymb,
preprint,%
]{revtex4-1}

\usepackage{graphicx}
\usepackage{dcolumn}
\usepackage{bm}
\usepackage{xcolor}
\usepackage{physics}
 \usepackage{multirow} 

\begin{document}

\preprint{AIP/123-QED}

\title{ \textcolor{black}{An Open Quantum System Theory for Polarizable Continuum Models}}
\author{Ciro A. Guido}

 \affiliation{Dipartimento di Scienze Chimiche, Universit\`a di Padova, Padova, Italy.}
 
 \author{Marta Rosa}
\affiliation{Dipartimento di Scienze Chimiche, Universit\`a di Padova, Padova, Italy.}
 
 \author{Roberto Cammi}

 \affiliation{Dipartimento di Chimica, Scienze della Vita e Sostenibilit\`a Ambientale, Universit\`a di Parma, Parma, Italy.}
 
 \author{Stefano Corni}
\email{stefano.corni@unipd.it}

 \affiliation{Dipartimento di Scienze Chimiche, Universit\`a di Padova, Padova, Italy.}
 \affiliation{CNR Istituto Nanoscienze, Modena, Italy}
 
\date{\today}

\begin{abstract}
{\color{black}{The problem of a solute described by Quantum Chemistry within a solvent represented as a polarizable continuum \textcolor{black}{model} (PCM) is here reformulated in terms of the open quantum systems (OQS) theory. Using its stochastic Schr\"{o}dinger Equation formulation, we are able to provide a more comprehensive picture of the electronic energies and of the coupling between solute and solvent electronic dynamics. In particular, OQS-PCM proves to be a unifying theoretical framework naturally including polarization and dispersion interactions, the effect of solvent fluctuations, and the non-Markovian solvent response. As such, the OQS-PCM describes the interplay between the solute and the solvent typical electronic dynamical times, and yields both the standard PCM and the so-called Born Oppenheimer solvation regime, where the solvent electronic response is considered faster than any electronic dynamics taking place in the solute. In analyzing the OQS-PCM, we obtained an expression for the solute-solvent dispersion (van der Waals) interactions that is very transparent in terms of a physical interpretation based on fluctuations and response functions. Finally, we present various numerical tests that support the theoretical findings
}}

\end{abstract}

\maketitle

\section{Introduction}
\textcolor{black}{The largest part of the photophysical and photochemical processes of technological and biological interest  takes place in condensed phase, or in presence of an environment surrounding the molecular system. It is therefore quite obvious that the theoretical investigation of molecules in solution is a central topic for the proper understanding of these molecular processes. Indeed, the 
environment may affect the electronic structure and the dynamics of the molecular system. In these cases, one of the most popular strategies in the computational Chemistry and Physics communities is the use of multiscale approaches:\cite{Warshel76,Field90,Thiel09,Tomasi2005,OniomReview,Curutchet2009,Olsen2010,Mennucci2019}  
the total system is partitioned in a active region (ranging from a single molecule to a supramolecular aggregate) which is 
the main responsible of the the observed process, property or signal, and an embedding environment that does not take a direct part in the process, but rather acts as a perturbation, thus affecting the behavior of the system. The first region is treated by a high level Quantum Chemistry method, whereas the rest is described by a lower level approach.} 
The two components can mutually interact but they typically keep their number of nuclei and electrons constant during the process investigated.  \\
Concerning the definition of the interactions between two subsystems, a hierarchy of  mechanisms can be pinpointed on the basis of their importance with respect to the process of interest. These mechanisms include general terms such as electrostatic, polarization, and dispersion as well as specific interactions such as hydrogen bonding. A multiscale approach to the description of molecular processes in the condensed phase should be able to account for all these mechanisms, or at least the most relevant ones depending on the specific process of interest.  This is quite a challenging task, since the network of interactions that determine the behavior of the system and its response properties can be rather complex. 
In this work, we will focus in particular on those models \cite{Tomasi2005,art:revklamt,art:revcramer} that combine a quantum-mechanical (QM) description of the active region with a representation of the environment as a structureless dielectric medium, mainly described by its macroscopic dielectric function, $\epsilon$, which determines the environment polarization as a response to the presence of the quantum system. 
In the most widespread formulations of these models (such as the polarizable continuum model (PCM) )\cite{Tomasi2005}
 the solvent polarization field is characterized by its value at the boundary between the molecule and the solvent medium. \cite{ASCCammiTomasi}   
 \\One of the focus here is the description of the polarization response of the solvent following an electronic excitation of the solute, and the proper coupling with the system's electronic structure: 
in continuum models, the study of electronic transitions
introduces new dimensions in the investigation of the problem.\cite{guido2019} One needs to describe
a correct time-dependent solvation regime: indeed, when a solute charge density is evolving in time, e.g., due to an external perturbation, the environment reaction field also becomes time-dependent in a non-trivial way due to the delayed response of the solvent polarization rooted in its frequency-dependent dielectric function.\cite{Cammi1995,cammibook,Corni2014} When the specific problem of calculating excited state energies (that can be thought as the result of a dynamical process prompted by an external perturbation) is investigated, a further element to be considered in continuum models is the scheme used to describe excited states. There are essentially two schemes in the framework of the standard PCM methods:\cite{Cammi2005} the state-specific (PCM-SS) approaches, which require the explicit calculation of the excited state wave-function, and the linear response (PCM-LR) descriptions that introduce effects related to the transition density only. Even if in isolated systems the PCM-SS and PCM-LR approaches are equivalent, assuming that the corresponding equations are solved exactly, in the presence of a polarizable environment,
the two formalisms give rise to a complementary description: the PCM-SS approach explicitly accounts for the rearrangement of the environment electronic degrees of freedom to adapt to the system excited state wavefunction, while the PCM-LR approach includes the term due to the dynamical response of the environment to the system charge density oscillating at the Bohr frequency. 
 The reason of such discrepancy is mainly due to the presence of the non-linear polarizable reaction field operator in the effective Hamiltonian.\cite{Corni2005}\\
 \textcolor{black}{There is also another point that has received less attention in the recent past, but that it is still not settled: standard PCM adopts 
 the assumption that the solute electrons interact with the solvent as an electron density cloud, i.e., it is assumed that the solvent response (both the nuclear and the electronic) is slow compared to the time scale of the solute electronic response.} This has been called the self-consistent field regime.\cite{Hynes92} The opposite situation, where the solvent electronic response is much faster than the solute electrons, is called Born-Oppenheimer regime and was explored by different authors with the specific problem of electron transfer in mind.\cite{Gehlen1992,Marcus1992,Hynes92,Basilevsky1994} 
In fact, the small value of the electronic excitation energy at the crossing of the barrier between reactants and products of the electron transfer reaction makes the dynamics of such transferring electron certainly slow compared to solvent electrons. \textcolor{black}{Since the first studies on the subject, \cite{Gehlen1992,Marcus1992} the inadequacy of a self consistent field regime in this situation was clear. A theory that bridges the gap between these two limits is needed. It can be also the starting point for elaborating controlled approximations for the intermediate regime.\cite{Marcus1992}}
 
To generalize the scope of quantum mechanical continuum solvation models, in particular for what regards the coupled solute-solvent electronic dynamics and the related problem of calculating excitation energies, we have started a research program to reformulate them, and specifically PCM, on the basis of the theory of open quantum systems (OQS).\cite{BreuerPetruccione}
Indeed, the coupling of a quantum system with the environment qualifies the probed system as open. The theory of the OQS is a fundamental approach to the understanding of the quantum dynamics of a microscopic system interacting with an environment (a thermal bath). In particular, one would like to neglect the intrinsic microscopic details of the latter, summarizing them in proper correlation functions and response functions, and to derive time dependent evolution equations (master equations) for the reduced density matrix of the system. 
\textcolor{black}{Even if the similarities between the basic idea of PCM and OQS are evident, the reformulation of PCM as a special case of OQS (with system $\rightarrow$ solute and bath $\rightarrow$ solvent) is far from obvious. This is mainly due to the different language normally used in the two approaches: i.e., the density matrix formalism for the system in OQS and that of wave-function for the solute in PCM.} 
However, within OQS theory, besides the conventional density matrix formalism, there has been an increasing interest over the years in the stochastic wave-function methods \cite{Breuer99,Gaspard1999,biele2012}, where the state of the open system is described by an ensemble of system wavefunctions, instead of a reduced density matrix. Such class of OQS approaches does not only provide a practical computational approach to investigate the electronic dynamics of molecules,\cite{Coccia2018} actually they are much closer to the PCM languange, and thus offer a good starting point to our goal.  We shall see that OQS provides a unifying framework for PCM, allowing to account for the solvent fluctuations and its non-Markovian and dissipative response effects on the solute electronic dynamics; bridging naturally between the self-consistent field (SCF) and the Born-Oppenheimer (BO) regimes and yielding estimates of solute excitation energies that includes both polarization and dispersion effects.  
Besides the formal derivation and the discussion of the theoretical results, in this work we also provide an initial numerical investigation of the resulting excitation energies for a few test cases: acrolein, methylencyclopropene (MCP) and para-nitroaniline (PNA) in two solvents  (dioxane and acetonitrile) of different polarity. \textcolor{black}{The goal of such numerical investigation is to illustrate the theoretical results.}

The paper is organized as follows. In the Theory section, we first provide a short introduction to PCM (sec. \ref{sec:PCM}) and to the OQS theory based on stochastic Schr\"odinger equations (sec. \ref{sec:OQS}). Then, starting from such theory we derive in secs.\ref{sec:choice}-\ref{sec:OQS_PCM_TD} the effective time-dependent Schr\"odinger equation for the solute wavefunction in a solvent described in the PCM fashion. We call such equation OQS-PCM time dependent Schr\"odinger equation. From such result, we move on to solute stationary states, leading to a OQS-PCM time independent  Schr\"odinger equation (sec. \ref{sec:OQS_PCM_TI}). Based on the latter, in sec. \ref{sec:SS} we discuss the expression of solvation energies for the ground and the excited states (and thus solvatochromic shifts) comparing them with standard PCM results, as well as with the recent GW/BSE PCM implementation.\cite{Duchemin2016,Duchemin2018} Then, in sec. \ref{sec:BO}, we describe how to obtain the BO regime out of the OQS-PCM, and related intermediate approximations.\cite{Marcus1992} At this stage of the discussion, we present (in sec. \ref{sec:disp}) an expression for the dispersion interaction energy that is written in terms of the same ingredients appearing in the OQS-PCM equations. We therefore show how dispersion is accounted for in OQS-PCM and in the BO limit. Finally, the Numerical section \ref{sec:numerical} and the Conclusions (sec. \ref{sec:conclusions}) follow.

\section{Theory}
\subsection{Foundation of the Open Quantum System Polarizable Continuum Model}
\subsubsection{Polarizable Continuum Model: basic definitions}
\label{sec:PCM}
We start this theoretical section by briefly summarizing the basics of PCM, to provide a consistent reference to concepts and nomenclature to the reminder of this work.

PCM is the standard approach, de facto, to perform quantum chemistry calculations of molecules in a polarizable solvent\cite{Tomasi2005,cammi1998calculation,cammi2000calculation,tomasi2002molecular,Cammi2005}. The latter is considered as a continuum and infinite dielectric medium characterized by a frequency-dependent dielectric function $\epsilon(\omega)$. The molecule,  described through quantum mechanics, is hosted in a molecular-shaped cavity 
\textcolor{black}{surrounded by} the otherwise homogeneous dielectric environment. The environment polarization, induced by the molecular charge density itself, is represented by a surface charge density, $\sigma_S$, spread over the surface of the molecular cavity containing the QM system, and obtained by solving a classical Poisson problem. Such problem is discretized by partitioning the cavity surface into finite elements: \textcolor{black}{during the years, multiple approaches have been proposed to discretize the cavity surface (see for instance refs.\citenum{Tomasi2005} and \citenum{Scalmani10} and references therein);} here we call such surface finite elements \emph{tesserae}, and we will represent the charge density as a collection of partial charges (apparent surface charges $\mathbf{q}$), placed in the tesserae centers.

The  time dependent PCM Schr\"odinger equation for the solute has been obtained through a variational formulation that includes non-equilibrium effects \cite{Cammi1995}, leading to a non-linear Schr\"odinger equation:
\begin{equation}\label{eq:sh_pcm}
{\it i}\frac{\partial|\Psi(t)\rangle}{\partial t} = [ \hat H_0 + \sum_{i} q_i[\Psi(t'<t)] \hat V_i + \hat {H}_{ext}(t)]|\Psi(t) \rangle
\end{equation}

where \textcolor{black}{atomic units have been used (as done throughout the text). In eq.(\ref{eq:sh_pcm}),} $\hat H_{0}$ is the Hamiltonian of the solute, $\boldsymbol{q}[\psi(t'<t)]$ represents the time dependent solvent polarization charges induced by the solute on the solvent (\textcolor{black}{which generally} depend on the entire history of the solute wavefunction $\psi(t')$ up to time $t$), $\boldsymbol{\hat V}$ is the molecular electrostatic potential of the solute at the tesserae centers;\cite{Tomasi2005,cammi1998calculation}  and $\hat {H}_{ext}(t)$ takes into account the solute interaction with an external time dependent perturbation (e.g., an electromagnetic field). The polarization charges $\boldsymbol{q}[\Psi(t'<t)]$ at time $t$ can be obtained as:\cite{Corni2014}

\begin{equation}\label{eq:q_pcm}
q_i[\Psi(t'<t)] = \int_{-\infty}^{t} \sum_j Q_{ij} (t - t') \langle\Psi(t') | \hat V_j | \Psi(t')\rangle dt'
\end{equation}
\noindent where ${\bf Q} (t - t')$ is the solvent response matrix, non-local in time and depending on the whole spectrum of the frequency-dependent dielectric permittivity of the medium. Eq.(\ref{eq:sh_pcm}) can therefore be rewritten as:
\begin{equation}\label{eq:sh_pcm_2}
{\it i}\frac{\partial|\Psi(t)\rangle}{\partial t} = [ \hat H_0 + \int_{-\infty}^{t} dt'\sum_{ij} Q_{ij} (t - t') \langle\Psi(t') | \hat V_j | \Psi(t')\rangle \hat V_i  + \hat {H}_{ext}(t)]|\Psi(t) \rangle
\end{equation}
which will be useful for comparison later in this work.

In the absence of external perturbations, stationary states for the solute can be obtained by solving the corresponding time independent PCM Schr\"odinger equation:

\begin{eqnarray}
\hat {H}^{PCM}_{\Phi'_A}|\Phi'_A\rangle &=& E^{PCM}_A|\Phi'_A\rangle 
\label{eq:TI_PCM}
\end{eqnarray}

where the effective PCM Hamiltonian written for the solute state $|\Phi'_A\rangle$  is given by:
\begin{eqnarray}
\hat {H}^{PCM}_{\Phi'_A}=\hat H_0 + \sum_{ij} Q_{0,ij} \langle \Phi'_A| \hat V_j | \Phi'_A \rangle  \hat V_i
\label{eq:H_PCM}
\end{eqnarray}

Eq.(\ref{eq:TI_PCM}) is a non-linear Schr\"odinger equation, since the effective Hamiltonian $\hat {H}^{PCM}_{\Phi'_A}$ depends on the wavefunction itself, and needs to be solved self-consistently for each target state. Moreover, the energy $E^{PCM}_A$ does not include the (free) energy cost of polarizing the solvent. When the solvent is fully equilibrated with the solute state, the proper (free) energy $\mathcal{G}_{sol}[\Phi'_A]$ that also includes this term is given by:
\begin{eqnarray}
\mathcal{G}_{sol}[\Phi'_A]=\langle \Phi'_A | \hat H_0 | \Phi'_A \rangle + \frac{1}{2} \sum_{ij} Q_{0,ij} \langle \Phi'_A| \hat V_j | \Phi'_A \rangle  \langle \Phi'_A |\hat V_i | \Phi'_A \rangle
\label{eq:G_PCM}
\end{eqnarray}

\subsubsection{Open quantum system theory: basic definitions and general form of the Stochastic Schr\"odinger equation}
\label{sec:OQS}
We want to reformulate the problem of a molecule within a PCM solvent in terms of the theory of OQS. \textcolor{black}{\cite{BreuerPetruccione}} 
The molecule is the quantum system whereas the \textcolor{black}{solvent} 
is the  \textcolor{black}{polarizable bath}. 
We are not assuming the Markovian approximation, rather we want to express the time-delayed response of the environment and its fluctuations in terms of the time dependent solvent polarization.
The Hamiltonian of the entire system, $\hat{H}_{tot}$ (molecule+environment, i.e., solute+solvent in this case) can be written as:
\begin{eqnarray}
\nonumber \hat{H}_{tot} & = & \hat{H}_{s}(t)  \otimes \hat{I}_{b} + \hat{I}_{s} \otimes \hat{H}_{b} + \lambda \hat{H}_{sb} \\ 
 \hat{H}_{sb} &=& \sum_{\alpha\beta} \hat{S}_{\alpha} \otimes \hat{B}_{\beta},
 \label{eq:start}
\end{eqnarray}

where $\hat{H}_s$ is the Hamiltonian of the molecular system, $\hat{H}_b$  the Hamiltonian of the the environment, \textcolor{black}{$\hat{I}_s$ and $\hat{I}_b$ are the identity operators acting on the systems and bath subspaces respectively}, $\hat{S}_\alpha$ is the operator acting on the molecular state and  $\hat{B}_{\beta}$ is the corresponding operator for the environment. Together they expresses the $\alpha\beta$-th interaction channel between the molecule and the environment , and $\lambda$ is a parameter to switch on the coupling ($\hat{H}_{sb}$), \textcolor{black}{hereafter set to be 1.}
\\
  Starting from eq.(\ref{eq:start}), Gaspard and Nagaoka\cite{Gaspard1999} have derived a non-Markovian stochastic Schr\"odinger Equation (SSE) for the system wavefunction (here the solute wavefunction) including a stochastic and a 
\textcolor{black}{history dependent} term for the partition given in eq.(\ref{eq:start}):

\begin{eqnarray}
\nonumber i\frac{\partial}{\partial t}|\Psi(t)\rangle &=&\hat{H}_s|\Psi(t)\rangle + \sum_{\alpha} \eta_{\alpha}(t)\hat{S}_\alpha |\Psi(t)\rangle +\\
&&-i  \int_{-\infty}^{t} dt' \sum_{\alpha \beta} C_{\alpha \beta}(t-t') \hat{S}_\alpha e^{-i\hat{H}_s (t-t')}\hat{S}_\beta |\Psi(t')\rangle
\label{eq:SSE}
\end{eqnarray}
 where atomic units have been used (as done through the text). In this equation, $\eta_{\alpha}(t)$ is a complex number representing the noise associated with the environment fluctuations,\cite{Gaspard1999} and satisfying:
\begin{eqnarray}
\overline{\eta_{\alpha}(t)}&=&0 \label{eq:noise}\\
\overline{\eta_{\alpha}(t)\eta_{\beta}(t')}&=&0\\
\overline{\eta_{\alpha}^*(t)\eta_{\beta}(t')}&=&C_{\alpha \beta}(t-t')=C_{ \beta \alpha}^*(t-t')
\end{eqnarray}
$C_{\alpha \beta}(t-t')$ is the correlation function of the environment, defined as:
\begin{eqnarray}
C_{\alpha \beta}(t-t')=\text{Tr}[\rho^{eq}_{env}\hat{B}^I_{\alpha}(t)\hat{B}^I_{\beta}(t')],
\end{eqnarray}
where $\rho^{eq}_{env}$ is the equilibrium density matrix of the environment (or bath) and $\hat{B}^I_{\alpha}(t)$ is the operator $\hat{B}$ of eq.(\ref{eq:start}) in the interaction picture (in the following we shall use the Schr\"odinger picture, but here the interaction picture makes the correlation function particularly transparent). In short, the evolution of the isolated system is modified by a stochastic term, representing the effect of the bath fluctuations (quantum and classical) plus a \textcolor{black}{dissipative} term, representing the effect of the response of the bath to the system evolution.

\subsubsection{ OQS-PCM: choice of the reference solute and solvent states}
\label{sec:choice}
{\color{black}{The first step toward reformulating PCM from open quantum system theory is to fix the unperturbed Hamiltonians in eq.(\ref{eq:start}) and the corresponding unperturbed states. To keep the theory and the \textcolor{black}{equations} 
as simple as possible, we define $\hat{H}_s$ to be the electronic Hamiltonian of the molecule in gas-phase, and $\hat{H}_b$ to be the electronic Hamiltonian of the unpolarized solvent, once the solute cavity has been built in it. \textcolor{black}{Notice that, for both solute and solvent molecules, we assume QM electrons while treating classically the nuclei.}
The two time independent Schr\"odinger equations
\begin{eqnarray}
\hat{H}_s | \Phi_I \rangle&=& E_I | \Phi_I \rangle \label{eq:unp_mol}\\
\hat{H}_b | \Upsilon_P \rangle&=& E_P | \Upsilon_P \rangle
\end{eqnarray}
 define the eigenstates $| \Phi_I \rangle $, $| \Upsilon_P \rangle$ and the corresponding eigenenergies $E_I$, $E_P$ of $\hat{H}_s$ and $\hat{H}_b$, respectively: for instance, $| \Phi_0 \rangle $ ($| \Upsilon_0 \rangle$) represents the solute (solvent) electronic ground state in the absence of the solvent (solute).} 
}\\

This choice for the solvent states means that we shall focus on the electronic response of the solvent only, i.e., the electron degrees of freedom of the solvent molecules get polarized by the solute, but the solvent molecules do not reorient. This is a suitable description for apolar solvents. In turn, the choice of the reference states for the molecule is made coherently: lacking an orientational polarization of the solvent, the electronic states of the solute are not initially polarized (the polarization will be developed by the interaction term in eq.(\ref{eq:start})).  
The extension to treat \textcolor{black}{a solvent of generic polarity} is based on the use of states that are equilibrated for both  solute and solvent at their ground states and then keeping frozen the solvent ones,\cite{Cammi2005} but it complicates the formula without providing more insights. \textcolor{black}{These frozen reaction field states encompass gas-phase states as a special case.} 
In the appendix \ref{app:polar} we shall briefly provide such extension, that is useful for the numerical tests as well.

We end this section by recalling that for apolar solvent the static ($\epsilon_0$) and the dynamic ($\epsilon_d$) dielectric constants are identical in practice, which means that the corresponding PCM matrices $\mathbf{Q}_0$ and $\mathbf{Q}_d$ are also approximately equals. Therefore, in the following, we shall refer only to $\mathbf{Q}_0$.

\subsubsection{Derivation of the OQS-PCM Time Dependent Schr\"odinger equation} 
\label{sec:OQS_PCM_TD}
Here we shall derive a time-dependent Schr\"odinger equation for a molecule in an implicit PCM solvent starting from the OQS theory. To this end, the first task to perform is to specify the operators $\hat{S}_\alpha$ and $\hat{B}_\alpha$ in the case of molecules in solution. In the literature on continuum solvation models, the solute-solvent interaction term appearing inside the solute effective Hamiltonian, $\hat{H}_{int}$, has been written before as:\cite{Hynes92,Georgievskii1999}
\begin{eqnarray}
\hat{H}_{int}=\int d\vec{r} \hat{\vec{E}}(\vec{r})\cdot\hat{\vec{P}}(\vec{r})
\end{eqnarray}   
where $\hat{E}(\vec{r})$ is the operator representing the electric field from the solute and $\hat{\vec{P}}(\vec{r})$ is the polarization vector operator acting on the solvent states. One can manipulate such interaction term involving operators as done before for the corresponding physical quantities\cite{Cammi1995} to rewrite $\hat{H}_{int}$ as:
  \begin{eqnarray}
\hat{H}_{int}=\int_{\Omega} d\vec{s}~ \hat{V}(\vec{s})\hat{\sigma}(\vec{s}) 
\end{eqnarray}
where the integral is done on the boundary $\Omega$ of the solute cavity, $\hat{\sigma}(\vec{s})=-\vec{n}(\vec{s})\cdot\hat{\vec{P}}(\vec{s})$ is the surface charge density operator acting on the solvent states, that can be obtained as the dot product of the outgoing normal to the cavity surface $\vec{n}(\vec{s})$ and the polarization vector operator $\hat{\vec{P}}(\vec{s})$, and $\hat{V}(\vec{s})$ is the electrostatic potential operator acting on the solute states. By discretizing the surface into tesseras, the integral can be rewritten as a sum over all the $i$:
  \begin{eqnarray}
\hat{H}_{int}=\sum_i \hat{V}_i \hat{q}_i
\label{eq:int} 
\end{eqnarray}
where $\hat{V}_i$ is the operator of the solute electrostatic potential acting on the center $\vec{s}_i$ and $ \hat{q}_i$ is the apparent surface charge operator for the same tessera, acting on solvent states. The comparison of eq.(\ref{eq:int}) with eq.(\ref{eq:start}) led us to identify  $\hat{V}_{i}$ as the operator $\hat{S}_{\alpha}$ and $\hat{q}_{i}$ as $\hat{B}_{\beta}$. 

\textcolor{black}{The correlation function of the environment in this specific case is therefore rewritten as} 
\begin{eqnarray}
C_{ij}(t-t')=\text{Tr}[\rho^{eq}_{sol}\hat{q}_{i}^I(t)\hat{q}^I_{j}(t')],
\end{eqnarray}
\textcolor{black}{where we have used again the interaction picture for the operators $\hat{q}_i$ and $\hat{q}_j$ as this makes the definition of the correlation function quite compact. Such correlation function can be written as\cite{Cohen04}}
\begin{eqnarray}\label{eq:c_cohen}
\nonumber C_{ij} (t-t')=C_{ij}^{sym} (t-t')+\frac{i}{2} \chi_{ij}(t-t') ~ ~for ~ ~(t-t')>0,\\ 
C_{ij} (t-t')=C_{ij}^{sym} (t-t')-\frac{i}{2} \chi_{ij}(t'-t) ~ ~for ~ ~(t-t')<0,
\end{eqnarray}
where $C_{ij}^{sym} (t-t')$ is the symmetric correlation function of the environment and $\chi_{ij}(t-t')$ is its linear susceptibility. Note the change of sign in the linear susceptibility with respect to ref.\citenum{Cohen04} since there a perturbation of the form $-\hat{S}\otimes \hat{B}$ (i.e., with a minus sign) was considered. Since $C_{ij}(t-t')$ is the 
correlation function \textcolor{black}{between apparent charges}, the meaning of the corresponding linear response function $\chi_{ij}(t-t')$ is clear: it gives the \textcolor{black}{value of the} apparent charge $q_{i}$ on the tessera $i$ once a given solute potential $V_{j}$ is applied to the tessera $j$. This is exactly the meaning of the time dependent PCM response matrix $Q_{ij}(t-t')$ in eq.(\ref{eq:q_pcm}),\textcolor{black}{\cite{Corni2014}} and we can therefore replace $\chi_{ij}(t'-t)$ with $Q_{ij}(t-t')$.

\textcolor{black}{The final form of SSE  is therefore obtained \textcolor{black}{by} replacing eq.(\ref{eq:c_cohen}) in eq.(\ref{eq:SSE})}
\begin{eqnarray}
\nonumber i\frac{\partial}{\partial t}|\Psi(t)\rangle &=&\hat{H}_s|\Psi(t)\rangle + \sum_{i} \eta_{i}(t)\hat{V}_i |\Psi(t)\rangle +\\
\nonumber &&-i \int_{-\infty}^{t} dt' \sum_{ij } C_{ij}^{sym} (t-t') \hat{V}_i e^{-i\hat{H}_s (t-t')}\hat{V}_j|\Psi(t')\rangle + \\
&+& \frac{1}{2} \int_{-\infty}^{t} dt' \sum_{ij }Q_{ij}(t-t') \hat{V}_i e^{-i\hat{H}_s (t-t')}\hat{V}_j|\Psi(t')\rangle.
\label{eq:SSE2}
\end{eqnarray}
This is the OQS-PCM time dependent  Schr\"odinger equation: and it is the first important result of this work. It represents a time dependent equation for the solute states in a PCM solvent including properly the concurrent dynamics of solute and solvent, and the effects of their fluctuations. In fact, by comparing with eq.(\ref{eq:sh_pcm}), we see that eq.(\ref{eq:SSE2}) includes two terms related to solvent fluctuations (the stochastic term involving $\eta_{i}(t)$ and the term involving the solvent symmetric correlation function $C_{ij}^{sym}$). Moreover, the term involving the PCM response matrix is more complex than in eq.(\ref{eq:sh_pcm}), \textcolor{black}{since} in eq.(\ref{eq:sh_pcm}) the relative solute-solvent time scales are \textcolor{black}{fixed} 
(in particular, the solute electrons are considered much faster than the solvent response, SCF regime).

Once a proper strategy to numerically evaluate the non-Markovian fluctuation terms of eq.(\ref{eq:SSE2}) is devised, such equation could be used to directly investigate the excitation process of a molecule in solution, providing a close theoretical counterpart to the experimental process without including the additional conceptual steps leading to state specific or linear response descriptions.\cite{guido2019} The direct numerical solution of this equation requires however further developments that are outside the scope of the present work.

\subsubsection{ Derivation of the Time Independent OQS-PCM Schr\"odinger equation}
\label{sec:OQS_PCM_TI}
In this section, we start from eq.(\ref{eq:SSE2}) to obtain an equation for stationary states that will provide an effective Hamiltonian to be compared directly with standard PCM results.  
In the spirit of PCM, the solvent fluctuations are averaged out from such equation, and we can neglect the effect of $\eta_{i}(t)\hat{V}_i$, as it follows from eq.(\ref{eq:noise}), to obtain:
\textcolor{black}{
\begin{eqnarray}\label{eq:SSEPCM}
\nonumber i\frac{\partial}{\partial t}|\Psi(t)\rangle &=&\hat{H}_s|\Psi(t)\rangle \\
 &&-i \int_{-\infty}^{t} dt' \sum_{ij } C_{ij}^{sym} (t-t') \hat{V}_i e^{-i\hat{H}_s (t-t')}\hat{V}_j|\Psi(t')\rangle + \\\nonumber
&+& \frac{1}{2} \int_{-\infty}^{t} dt' \sum_{ij }Q_{ij}(t-t') \hat{V}_i e^{-i\hat{H}_s (t-t')}\hat{V}_j|\Psi(t')\rangle.
\end{eqnarray}
}
Assuming a stationary solution of eq.(\ref{eq:SSEPCM}) in the form, $|\Psi(t)\rangle=exp(-iE'_At)|\Phi'_A\rangle$ we get:

 \begin{eqnarray}
\nonumber E'_A|\Phi'_A\rangle &=&\hat{H}_s|\Phi'_A\rangle +\\
\nonumber &&-i \int_{-\infty}^{t} dt' e^{-iE'_A(t'-t)}\sum_{ij } C_{ij}^{sym} (t-t') \hat{V}_i e^{-i\hat{H}_s (t-t')}\hat{V}_j|\Phi'_A\rangle + \\
&+& \frac{1}{2} \int_{-\infty}^{t} dt' e^{-iE'_A(t'-t)}\sum_{ij}Q_{ij}(t-t') \hat{V}_i e^{-i\hat{H}_s (t-t')}\hat{V}_j|\Phi'_A\rangle.
\label{eq:SSEs}
\end{eqnarray}

\textcolor{black}{We recall that here the reference state is the unperturbed state for both the solute and the solvent. This expression can be reworked by inserting} the resolution of identity $\sum_K |\Phi_K\rangle\langle \Phi_K | $ in the last two terms (where $|\Phi_K\rangle$ are the eigenstates of $\hat{H}_s$ with eigenvalues $E_K$ as defined in eq.(\ref{eq:unp_mol})), just before the operator $\hat{V}_j$:
\begin{eqnarray}
\nonumber E'_A|\Phi'_A\rangle &=&\hat{H}_s|\Phi'_A\rangle +\\
\nonumber &&-i \sum_K \int_{-\infty}^{t} dt' e^{-i\omega_{A'K}(t'-t)}\sum_{ij } C_{ij}^{sym} (t-t') \hat{V}_i |\Phi_K\rangle\langle \Phi_K |\hat{V}_j|\Phi'_A\rangle + \\
&+& \frac{1}{2} \sum_K \int_{-\infty}^{t} dt' e^{-i\omega_{A'K}(t'-t)}Q_{ij}(t-t') \hat{V}_i |\Phi_K\rangle\langle \Phi_K | \hat{V}_j |\Phi'_A\rangle.
\label{eq:SSEs2}
\end{eqnarray}

where the frequencies $\omega_{A'K}=E'_A-E_K$ have been introduced. To simplify the integrals, we also introduce $\tau=t-t'$ to get:
\begin{eqnarray}
\nonumber E'_A|\Phi'_A\rangle &=&\hat{H}_s|\Phi'_A\rangle +\\
\nonumber &&-i \sum_K \int_{0}^{\infty} d\tau e^{i\omega_{A'K}\tau}\sum_{ij } C_{ij}^{sym} (\tau) \hat{V}_i |\Phi_K\rangle\langle \Phi_K |\hat{V}_j|\Phi'_A\rangle + \\
&+& \frac{1}{2} \sum_K \int_{0}^{\infty} d\tau e^{i\omega_{A'K}\tau}\sum_{ij}Q_{ij}(\tau) \hat{V}_i |\Phi_K\rangle\langle \Phi_K | \hat{V}_j|\Phi'_A\rangle.
\label{eq:SSEk}
\end{eqnarray}

The integrals in eq.(\ref{eq:SSEk}) can be performed in terms of the Fourier transform of $C_{ij}^{sym} (\tau)$ and $Q_{ij}(\tau)$. Indeed, recalling that $C_{ij}^{sym} (\tau)$ is an even function of $\tau$ and that $Q_{ij}(\tau)=0$ when $\tau<0$ we get:
\begin{eqnarray}
\int_{0}^{\infty} d\tau e^{i\omega_{A'K}\tau} C_{ij}^{sym} (\tau)=\frac{1}{2}C_{ij}^{sym}(\omega_{A'K})\\
\int_{0}^{\infty} d\tau e^{i\omega_{A'K}\tau} Q_{ij} (\tau)=Q_{ij}(\omega_{A'K})
\end{eqnarray}
Therefore,
\begin{eqnarray}
\nonumber E'_A|\Phi'_A\rangle &=&\hat{H}_s|\Phi'_A\rangle +\\
\nonumber &&-\frac{i}{2} \sum_K \sum_{ij } C_{ij}^{sym} (\omega_{A'K}) \hat{V}_i |\Phi_K\rangle\langle \Phi_K |\hat{V}_j|\Phi'_A\rangle + \\
&+& \frac{1}{2} \sum_K \sum_{ij}Q_{ij}(\omega_{A'K}) \hat{V}_i |\Phi_K\rangle\langle \Phi_K | \hat{V}_j|\Phi'_A\rangle.
\label{eq:SSEs3}
\end{eqnarray}
\\
We have now to \textcolor{black}{rewrite} $C_{ij}^{sym} (\omega_{A'K}) $ in terms of standard PCM quantities. {\color{black}{To this end, we take advantage of the fluctuation-dissipation theorem to connect the symmetric correlation function $C_{ij}^{sym} (\omega_{A'K}) $ to the imaginary part, $Q_{ij}''(\omega_{A'K})$, of the linear response of the polarizable bath $Q_{ij}(\omega_{A'K})$ \cite{Cohen04}:
\begin{eqnarray}\label{eq:c_sym}
C_{ij}^{sym} (\omega_{A'K})=- \coth{(\frac{\omega_{A'K}}{2kT})}Q_{ij}''(\omega_{A'K})
\end{eqnarray}
 \textcolor{black}{It should be noted} the change of sign with respect to ref.\citenum{Cohen04}, in line with \textcolor{black}{what} was already considered for eq(\ref{eq:c_cohen}).}}\\
 We can assume that either $\omega_{A'K}$ is a small quantity, that can be approximated with 0, when $|\Phi'_A\rangle$ is the solvent-perturbed version of $|\Phi_K\rangle$ (i.e. $|\Phi'_A\rangle = |\Phi'_K\rangle$), or $\omega_{A'K}$ is a quantity of the order of an electronic excitation energy (much larger than $k_BT$ at room temperature), when the two states are solvent-perturbed version of different states (i.e. $|\Phi'_A\rangle \neq |\Phi'_K\rangle$). In this low-temperature limit, eq.(\ref{eq:c_sym}) is reduced to 
\begin{eqnarray}
C_{ij}^{sym} (\omega_{A'K})=\Theta(-\omega_{A'K})Q_{ij}''(\omega_{A'K})-\Theta(\omega_{A'K})Q_{ij}''(\omega_{A'K})
\label{eq:fd}
\end{eqnarray}
\textcolor{black}{Here the step functions take into account the sign dependence from  $E'_A-E_K$ of hyperbolic cotangent function.}
As a consequence, eq.(\ref{eq:SSEs3}) is reduced to
\begin{eqnarray}
&&\hat{H}_s|\Phi'_A\rangle + \nonumber\\
\nonumber &+&\frac{i}{2} \sum_K \sum_{ij } (1+\Theta(\omega_{A'K})-\Theta(-\omega_{A'K})) Q_{ij}'' (\omega_{A'K}) \hat{V}_i |\Phi_K\rangle\langle \Phi_K |\hat{V}_j|\Phi'_A\rangle + \\
&+& \frac{1}{2} \sum_K \sum_{ij}Q_{ij}'(\omega_{A'K}) \hat{V}_i |\Phi_K\rangle\langle \Phi_K | \hat{V}_j|\Phi'_A\rangle= E'_A|\Phi'_A\rangle.
\label{eq:SSEnoC}
\end{eqnarray}
\textcolor{black}{where $Q_{ij}'(\omega_{A'K})$ is the real part of the linear response of the polarizable bath $Q_{ij}(\omega_{A'K})$.} Eq.(\ref{eq:SSEnoC}) is the time-independent OQS-PCM Schr\"odinger equation, which can be written in a compact form:
\begin{eqnarray}
\hat{H}^{OQS}_{\Phi'_A}|\Phi'_A\rangle &=&   E'_A|\Phi'_A\rangle
\label{eq:TI_SE}
\end{eqnarray}
by defining the effective, state-specific OQS-PCM Hamiltonian $\hat{H}^{OQS}_{\Phi'_A}$ as:
\begin{eqnarray}
\nonumber \hat{H}^{OQS}_{\Phi'_A} &=&\hat{H}_s +\\
\nonumber &+&\frac{i}{2} \sum_K \sum_{ij } (1+\Theta(\omega_{A'K})-\Theta(-\omega_{A'K})) Q_{ij}'' (\omega_{A'K}) \hat{V}_i |\Phi_K\rangle\langle \Phi_K |\hat{V}_j + \\
&+& \frac{1}{2} \sum_K \sum_{ij}Q_{ij}'(\omega_{A'K}) \hat{V}_i |\Phi_K\rangle\langle \Phi_K | \hat{V}_j
\label{eq:H_OQS}
\end{eqnarray}
State specificity \textcolor{black}{arises} by the presence of $E'_A$ inside the Hamiltonian expression, and not by the direct involvement of the target state $|\Phi'_A\rangle$ inside the Hamiltonian as it happens for standard PCM. It \textcolor{black}{is reminiscent}  of the Brillouin-Wigner perturbation theory,\cite{Magnasco2013} where the perturbative correction depends on the exact energy of the state.  It is also to be remarked that the OQS-PCM time independent Schr\"odinger equation contains the frequency-dependent PCM matrix $\mathbf{Q}(\omega)$, for values of $\omega$ that are related to excitation energies of the solute. On the contrary, the standard PCM equation contains only reference to the static PCM matrix $\mathbf{Q}_0$. Finally, the most obvious difference between the OQS-PCM vs PCM effective Hamiltonian is the presence of an imaginary part in the former, a consequence of the capability of the environment  to dissipate \textcolor{black}{the} energy of the \textcolor{black}{solute}. This will be further clarified in the next section.

To obtain an approximated expression for $E_A'$, i.e., the energy associated to the stationary states of the OQS-PCM effective Hamiltonian, we now express eq.(\ref{eq:TI_SE}) in a matrix form, by writing the state 
\textcolor{black}{$|\Phi'_A\rangle$} as a linear combination of the \textcolor{black}{solute} gas-phase states $|\Phi_J\rangle$:
\textcolor{black}{
\begin{eqnarray}
|\Phi'_A\rangle=\sum_J |\Phi_J\rangle C'_{JA}
\end{eqnarray}
}
By replacing such expression in eq.(\ref{eq:TI_SE}), and multiply the latter by $\langle \Phi_I |$, \textcolor{black}{we obtain}:
\begin{eqnarray}
\sum_J H_{\Phi'_A,IJ}^{OQS} C'_{JA}=E_A' C'_{IA}
\label{eq:TI}
\end{eqnarray}
where:
\begin{eqnarray}
\nonumber H_{\Phi'_A,IJ}^{OQS}&=&E_I\delta_{IJ} +\\
\nonumber &+&\frac{i}{2} \sum_K \sum_{ij } (1+\Theta(\omega_{A'K})-\Theta(-\omega_{A'K})) Q_{ij}'' (\omega_{A'K}) \langle\Phi_I|\hat{V}_i |\Phi_K\rangle\langle \Phi_K |\hat{V}_j|\Phi_J\rangle + \\
&+& \frac{1}{2} \sum_K \sum_{ij}Q_{ij}'(\omega_{A'K}) \langle\Phi_I| \hat{V}_i |\Phi_K\rangle\langle \Phi_K | \hat{V}_j| \Phi_J \rangle.
\label{eq:HIJ}
\end{eqnarray}

Eqs.(\ref{eq:TI_SE})-(\ref{eq:HIJ}) are the key results of this section, allowing to find stationary states and their energies in solution.
\textcolor{black}{In the next two sections,} 
we explore their implication in terms of solvation energy, compare them with standard PCM results and connect the various terms with standard mechanism of solute/solvent interaction such as polarization and dispersion. 

\subsection{\textcolor{black}{Solvation} energies from the OQS-PCM time-independent Schr\"odinger equation}
\label{sec:SS}
Considering the interaction with the solvent as a perturbation, the effect to the lowest non-zero order (the second) on the ground and excited state energies can be obtained by considering only the diagonal elements of $\mathbf{H}^{OQS}$ defined in eq.(\ref{eq:HIJ}), i.e., $E^{(2)}_{I,OQS}=H_{II}^{OQS}$ (out of diagonal element contribute to higher order). We are analyzing such energies in the next section, and we shall demonstrate that the physics included into them is in line with the standard state specific treatment of PCM. 

\subsubsection{Ground state energy}
\label{sec:gs}
\textcolor{black}{Let us first consider the ground state case.
If we assume $E'_0 \approx E_0$ in $H_{00}$, the GS energy corrected to the second order becomes}:
\begin{eqnarray}
\nonumber E^{(2)}_{0,OQS}&=&E_0 +\\
\nonumber &+&\frac{i}{2} \sum_K \sum_{ij } (1+\Theta(\omega_{0K})-\Theta(\omega_{K0})) Q_{ij}'' (\omega_{0K}) \langle\Phi_0|\hat{V}_i |\Phi_K\rangle\langle \Phi_K |\hat{V}_j|\Phi_0\rangle + \\
&+& \frac{1}{2} \sum_K \sum_{ij}Q_{ij}'(\omega_{0K}) \langle\Phi_0 | \hat{V}_i |\Phi_K\rangle\langle \Phi_K | \hat{V}_j| \Phi_0 \rangle.
\label{eq:H00}
\end{eqnarray}

\textcolor{black}{where $\omega_{K0}=E_K-E_0$ and $\omega_{0K}=E_0-E_K$. This expression can be simplified, separating} 
the sum in $K$ in the term with $K=0$ and those with $K\neq0$. 
\textcolor{black}{Indeed,} since $E_{K}-E_0>0$ for all the $K\neq0$, and since  $Q_{ij}'' (E_0-E_0)=Q_{0,ij}''=0$, the imaginary term in the r.h.s. of eq.(\ref{eq:H00}) is identically 0. We therefore get:
\begin{eqnarray}
\nonumber E^{(2)}_{0,OQS}&=&E_0 + \frac{1}{2} \sum_{ij}Q_{0,ij} \langle\Phi_0| \hat{V}_i |\Phi_0\rangle\langle \Phi_0 | \hat{V}_j| \Phi_0 \rangle\\
&+& \frac{1}{2} \sum_{K\neq0} \sum_{ij}Q_{ij}'(\omega_{K0}) \langle\Phi_0| \hat{V}_i |\Phi_K\rangle\langle \Phi_K | \hat{V}_j| \Phi_0 \rangle.
\label{eq:H00s}
\end{eqnarray}
\textcolor{black}{Here, we also exploited the property of $Q_{ij}'(\omega)$ to be a even function of $\omega$ to write $Q_{ij}'(\omega_{0K})=Q_{ij}'(\omega_{K0})$}.\\
\textcolor{black}{The final step is to recast eq.(\ref{eq:H00s}) in the usual PCM like form, using the static and the real part of the dynamic PCM response matrix to define the charges:
\begin{eqnarray}
\mathbf{q}^0&=&\mathbf{Q}_0 \langle\Phi_0 | \mathbf{\hat{V}} | \Phi_0 \rangle\\
\mathbf{q}^{0K}&=&\mathbf{Q}'(\omega_{K0}) \langle \Phi_0 | \mathbf{\hat{V}} | \Phi_K \rangle
\end{eqnarray}
to finally get:}
\begin{eqnarray}
\nonumber E^{(2)}_{0,OQS}&=&E_0 + \frac{1}{2} \sum_i q^{0}_i\langle \Phi_0 | \hat{V}_i | \Phi_0 \rangle\\
&+& \frac{1}{2} \sum_{K\neq0,i} q^{0K}_i\langle \Phi_K | \hat{V}_i| \Phi_0 \rangle.
\label{eq:H00sij}
\end{eqnarray}
Already at the level of ground state energy, $E^{(2)}_0$ is intriguing. First, the usual PCM solvation free energy (albeit at this lowest level of perturbation theory, i.e., based on in vacuo states) appears explicitly. Note that, being the reference state of the solvent the unpolarized one, \textcolor{black}{it appears} the proper factor $\frac{1}{2}$, \textcolor{black}{which} takes into account the energy needed to polarize the solvent. Secondly, an extra term \textcolor{black}{comes out}, that involves excited states and the solvent PCM response matrix evaluated at excitation energies. The meaning of such term will be clarified in sect. \ref{sec:disp}, where its relation with dispersion interactions will be discussed.

{\it Self Consistent field regime.} It is informative to see what happens to eq.(\ref{eq:H00s}), when the solvent electrons are deemed \textcolor{black}{to be} very slow compared to solute electrons, i.e., the so called SCF limit.\cite{Hynes92} In this limit, one can assume that $Q_{ij}'(\omega_{K0})$ is null for $K\neq0$, i.e., the solvent is not able to respond to any solute excitation, which are too fast for the solvent to follow. In such limit, the second line of eq.(\ref{eq:H00s}) is 0, and we are left with:
\begin{eqnarray}
E^{(2)}_{0,SCF}&=&E_0 + \frac{1}{2} \sum_i q^{0}_i\langle \Phi_0 | \hat{V}_i | \Phi_0 \rangle
\label{eq:H00SCF}
\end{eqnarray}
which is the usual PCM-SS expression (but here the gas-phase state is involved) including the electrostatic polarization. 

\subsubsection{Excited state energies}
\label{sec:es}
For a generic excited state $I>0$: 
\begin{eqnarray}
\nonumber E^{(2)}_{I,OQS}&=&E_I +\\
\nonumber &+&\frac{i}{2} \sum_K \sum_{ij } (1+\Theta(\omega_{IK})-\Theta(\omega_{KI})) Q_{ij}'' (\omega_{IK}) \langle\Phi_I|\hat{V}_i |\Phi_K\rangle\langle \Phi_K |\hat{V}_j|\Phi_I\rangle + \\
&+& \frac{1}{2} \sum_K \sum_{ij}Q_{ij}'(\omega_{IK}) \langle\Phi_I | \hat{V}_i |\Phi_K\rangle\langle \Phi_K | \hat{V}_j| \Phi_I \rangle.
\label{eq:HII}
\end{eqnarray}
As before, we can separate $K=I$ from all the other values of $K$ in the sums. However, in the second line of eq.(\ref{eq:HII}) we have now to take care of the sign of $E_I-E_K$: when it is negative ($E_K>E_I$), the term in parenthesis sums to zero, when it is positive ($E_K<E_I$) it sums to 2. We therefore get:
 \begin{eqnarray}
\nonumber E^{(2)}_{I,OQS}&=&E_I +\\
\nonumber &+&i \sum_{E_K<E_I} \sum_{ij }  Q_{ij}'' (\omega_{IK}) \langle\Phi_I|\hat{V}_i |\Phi_K\rangle\langle \Phi_K |\hat{V}_j|\Phi_I\rangle + \\
\nonumber &+& \frac{1}{2} \sum_{ij}Q_{0,ij} \langle\Phi_I | \hat{V}_i |\Phi_I\rangle\langle \Phi_I | \hat{V}_j| \Phi_I \rangle\\
&+& \frac{1}{2} \sum_{K\neq I} \sum_{ij}Q_{ij}'(\omega_{IK}) \langle\Phi_I | \hat{V}_i |\Phi_K\rangle\langle \Phi_K | \hat{V}_j| \Phi_I \rangle
\label{eq:HII2}
\end{eqnarray}
The most obvious difference with the ground state \textcolor{black}{result is that here the energy includes also an imaginary part. This means that} the population of the excited state $I$ is exponentially decreasing over time.\cite{Norman2005} The reason why this happens is quite clear: the state $I$ can decay to lower energy states $K$ by transferring the excess energy to the solvent, that can absorb it \textcolor{black}{with} a rate proportional to the dissipative (imaginary) part of its response functions $Q_{ij}'' (\omega_{IK}) $. The latter, based on the response function definition we are using here, is negative for $E_I-E_K>0$. \textcolor{black}{Obviously the GS cannot decay, being the lowest-energy state, and indeed eq.(\ref{eq:H00s}) contains just the real term}. The value of the excited state imaginary energy (doubled) provides an estimate of the excited state decay rate induced by energy transfer to the solvent, and thus \textcolor{black}{is of practical use}. In the following, we shall focus on the effects of the solvent on energies, and we shall therefore discuss these imaginary terms no further.
The second line of eq.(\ref{eq:HII2}) is the usual PCM like polarization energy in state $I$, while the last line provides a term (analogous to that already found for the ground state) that will be discussed later.

{\it Excitation energies.} Based on the results obtained so far, we can now write the excitation energy as:
 \begin{eqnarray}
\nonumber \Delta E^{(2)}_{I0,OQS}&=&\Delta E_{I0} +\\
\nonumber &+& \frac{1}{2} \sum_{ij}Q_{0,ij} \langle\Phi_I | \hat{V}_i |\Phi_I\rangle\langle \Phi_I | \hat{V}_j| \Phi_I \rangle\\
\nonumber &-& \frac{1}{2} \sum_{ij}Q_{0,ij} \langle\Phi_0 | \hat{V}_i |\Phi_0\rangle\langle \Phi_0 | \hat{V}_j| \Phi_0 \rangle\\
\nonumber &+& \frac{1}{2} \sum_{K\neq I} \sum_{ij}Q_{ij}'(\omega_{IK}) \langle\Phi_I | \hat{V}_i |\Phi_K\rangle\langle \Phi_K | \hat{V}_j| \Phi_I \rangle+\\
&-& \frac{1}{2} \sum_{K\neq 0,} \sum_{ij}Q_{ij}'(\omega_{0K}) \langle\Phi_0 | \hat{V}_i |\Phi_K\rangle\langle \Phi_K | \hat{V}_j| \Phi_0 \rangle
\label{eq:DE}
\end{eqnarray}

{\color{black}{This last equation can be rearranged,
adding and subtracting the quantity
\begin{equation}
\nonumber
\sum_{ij}Q_{0,ij}\langle\Phi_0 | \hat{V}_i |\Phi_0\rangle\langle \Phi_I | \hat{V}_j| \Phi_I \rangle
\end{equation}

and introducing the so-called frozen solvent excitation energy\cite{Cammi2005} $\Delta E_{I0}^{fro}$ 
\begin{eqnarray}\label{eq:frozen}
\Delta E_{I0}^{fro}=E_I-E_0 +\sum_{ij}Q_{0,ij}\langle\Phi_0 | \hat{V}_i |\Phi_0\rangle \left( \langle \Phi_I | \hat{V}_j| \Phi_I \rangle - \langle \Phi_0 | \hat{V}_j| \Phi_0\rangle \right)
\end{eqnarray}
to obtain the final form of the excitation energy:
 \begin{eqnarray}\label{eq:DE2}
\nonumber \Delta E^{(2)}_{I0,OQS}&=&\Delta E_{I0}^{fro} +\\
\nonumber &+& \frac{1}{2} \sum_{ij} \left( \langle\Phi_I | \hat{V}_i |\Phi_I\rangle-\langle \Phi_0 | \hat{V}_i| \Phi_0 \rangle \right) \left( \langle\Phi_I | Q_{0,ij}\hat{V}_j |\Phi_I\rangle-\langle \Phi_0 | Q_{0,ij}\hat{V}_j| \Phi_0 \rangle \right)\\
\nonumber &+& \frac{1}{2} \sum_{K\neq I} \sum_{ij}Q_{ij}'(\omega_{IK}) \langle\Phi_I | \hat{V}_i |\Phi_K\rangle\langle \Phi_K | \hat{V}_j| \Phi_I \rangle+\\
&-& \frac{1}{2} \sum_{K\neq 0} \sum_{ij}Q_{ij}'(\omega_{0K}) \langle\Phi_0 | \hat{V}_i |\Phi_K\rangle\langle \Phi_K | \hat{V}_j| \Phi_0 \rangle
\end{eqnarray}
}}

This excitation energy is again familiar, as it represents the state-specific expression (first and second lines) plus terms associated with dispersion (third and fourth lines), as it will be clarified in sect. \ref{sec:disp}. The same expression for a \textcolor{black}{a solvent of generic polarity, that make use of frozen-solvent solute states,} can be found in Appendix \ref{app:polar}. 

{\it Relation with PCM-LR excitation energies}. One point that deserves a comment is whether this form of the excitation energy also contains a linear response-like term, as found before.\cite{Corni2005} Such term is proportional to 
\textcolor{black}{the square of the $0$-$I$ reaction transition potential (i.e. the one induced by the $0$-$I$ transition density).}
The latter appears in the last two terms of the equation (\ref{eq:DE2}), taking $K=0$ in the third line and $K=I$ in the fourth line. Yet, we see that the two terms resulting from these choices have opposite signs, and therefore cancel each other. At this stage, the linear response term does not appear in the excitation energies. We shall see in sect. \ref{sec:dis_pcm} that this is a consequence of one of the assumption we made in the present derivation. 

{\it Relation with the GW/BSE-PCM excitation energies}. Finally, we would like to comment on the relation between the equation presented here and those obtained within the framework of GW/BSE.\cite{Duchemin2016,Duchemin2018} The single particle energies $\epsilon_n^{GW}$ provide estimates of $-IP_n$ (with $IP_n$ the ionization potential from the orbital $n$, i.e., $n=HOMO$ would be the first ionization potential) when $n$ is an occupied orbital, and of $-EA_n$ (the electron affinity adding an electron in the orbital $n$) when $n$ is a virtual one. The solvent-induced shift of the GW self energy (additional to the frozen solvent contribution) for the orbital $n$ in the COHSEX approximation (Coulomb-hole plus screened exchange), $\delta \epsilon_n^{GW}$ can be approximated as\cite{Duchemin2018} (using the present notation):
\begin{eqnarray}
\delta \epsilon_n^{GW}=\frac{1}{2}\sum_a^{virt}\sum_{ij}V_i^{na}Q_{0,ij}V_j^{an}-\frac{1}{2}\sum_m^{occ}\sum_{ij}V_i^{nm}Q_{0,ij}V_j^{mn}
\label{eq:GW}
\end{eqnarray}

where $V_i^{na}$ is the potential generated on the tessera $i$ by the charge distribution obtained by the product of the orbitals $\phi_n$ and $\phi_a$  (we consider real orbitals). The first sum in the r.h.s. runs on the {\it virt}ual orbitals, the second sum runs on the {\it occ}upied ones.  We shall now show that eq.(\ref{eq:GW}) is coherent with eq.(\ref{eq:DE2}). When $n$ is an occupied orbital, we have to compare $\delta \epsilon_n^{GW}$ with the solvent-induced shift (beyond the frozen solvent contribution) of $-IP_n$, that we can calculate from $-\Delta E^{(2)}_{I0,OQS}$ in eq.(\ref{eq:DE2}) choosing $|\Phi_I \rangle$ to be the ionized state and $|\Phi_0 \rangle$ the neutral state. In particular, the solvent-induced shift we are looking for (called here $\delta \epsilon_n^{OQS}$) is minus the sum of the last three lines of eq.(\ref{eq:DE2}):
 \begin{eqnarray}\label{eq:GW_OQS_d}
\nonumber \delta \epsilon_n^{OQS}&=&-\frac{1}{2} \sum_{ij} \left( \langle\Phi_I | \hat{V}_i |\Phi_I\rangle-\langle \Phi_0 | \hat{V}_i| \Phi_0 \rangle \right) \left( \langle\Phi_I | Q_{0,ij}\hat{V}_j |\Phi_I\rangle-\langle \Phi_0 | Q_{0,ij}\hat{V}_j| \Phi_0 \rangle \right)\\
\nonumber &-& \frac{1}{2} \sum_{K\neq I} \sum_{ij}Q_{ij}'(\omega_{IK}) \langle\Phi_I | \hat{V}_i |\Phi_K\rangle\langle \Phi_K | \hat{V}_j| \Phi_I \rangle+\\
&+& \frac{1}{2} \sum_{K\neq 0} \sum_{ij}Q_{ij}'(\omega_{0K}) \langle\Phi_0 | \hat{V}_i |\Phi_K\rangle\langle \Phi_K | \hat{V}_j| \Phi_0 \rangle
\end{eqnarray}

To show the equivalence of $\delta \epsilon_n^{GW}$ and $\delta \epsilon_n^{OQS}$, we have to make further assumptions. The states $|\Phi_K \rangle$ are approximated with single Slater determinants, and $|\Phi_I \rangle$ is the state of the molecule where an electron has been removed by the orbital $n$ occupied in $|\Phi_0 \rangle$ without allowing the other orbitals to relax. With this choice of states, we have that:
\begin{eqnarray}
\nonumber \delta \epsilon_n^{OQS}&=& - \frac{1}{2}\sum_{ij}V_i^{nn}Q_{0,ij}V_j^{nn}+\\
&-&\frac{1}{2}\sum_{m\neq n}^{occ}\sum_{ij}V_i^{mn}Q_{ij}'(\omega_mn)V_j^{nm} +\frac{1}{2} \sum_{a}^{virt}\sum_{ij}V_i^{na}Q_{ij}'(\omega_{na})V_j^{an}
\label{eq:GW_OQS}
\end{eqnarray}
which is equal to $\delta \epsilon_n^{GW}$ in eq.(\ref{eq:GW}) when we assume $Q_{ij}'(\omega_{mn})=Q_{ij}'(\omega_{na})=Q_{0,ij}$ as in refs. \onlinecite{Duchemin2016,Duchemin2018}. \textcolor{black}{This assumption, which will be discussed further in the next section,  leads to the so called Born-Oppenheimer (BO) approximation. The expression  for $\delta \epsilon_n^{GW}$ of eq.(\ref{eq:GW}), which is an approximation to what actually implemented in refs. \onlinecite{Duchemin2016,Duchemin2018}, has an accuracy beyond what would seem to imply the assumptions we did on the states $|\Phi_K \rangle$.} Yet, these simplifications allowed to show that the basic physics of OQS-PCM is included in the GW-PCM  (and thus in the GW/BSE-PCM) approach. 
  


\subsection{The OQS-PCM Hamiltonian in the Born-Oppenheimer \textcolor{black}{regime}}
\label{sec:BO}
Let us analyze here how the  the solvation energy expressions found so far change when one assumes that {\it all} the solute excitation energies are much smaller than any solvent excitation energy, 
i.e., that the electronic dynamics of the solvent is much faster than the solute one. This is the BO \textcolor{black}{regime}.\cite{Hynes92} In this limit, $ Q_{ij}'(\omega_{J'K})\approx Q_{0,ij}$, i.e., the solvent sees \textcolor{black}{every} excitation energy of the solute as very small. The term $Q_{0,ij}$ is independent from $K$ and thus can be taken out from the sum over $K$ in eq.(\ref{eq:HIJ}). By exploiting the resolution of the identity $\sum_K |\Phi_K\rangle\langle \Phi_K | =1 $ and simplifying the QOS-PCM Hamiltonian by neglecting its imaginary part  (that we have already connected to the decay of the excited states), we obtain:
\begin{eqnarray}
 H_{IJ}^{BO}&=&E_I\delta_{IJ} + \frac{1}{2} \sum_{ij} \langle\Phi_I| \hat{V}_i Q_{0,ij}  \hat{V}_j| \Phi_J \rangle
\label{eq:HIJBO}
\end{eqnarray}

The latter is essentially the expression for the solvent Born-Oppenheimer Hamiltonian proposed before by Basilevski et al. \cite{Basilevsky1994} As noted there, in this limit the solvent term in the Hamiltonian amounts to include the solvent dielectric screening in the electron-electron Coulomb interaction rather than on the electron solute cloud. This is readily understood if we write explicitly the operators $\hat{V}_i$ and $\hat{V}_j$ in terms of the $u$ and $v$ electron positions, $\vec{r}_u$ and $\vec{r}_v$, and the $i$ and $j$ tessera centers, $\vec{s}_i$ and $\vec{s}_j$:
 \begin{eqnarray}
H_{IJ}^{BO}&=&E_I\delta_{IJ} + \frac{1}{2} \sum_{uv,ij} \langle\Phi_I | \frac{1}{|\vec{r}_u-\vec{s}_i|} Q_{0,ij} \frac{1}{|\vec{s}_j-\vec{r}_v|} | \Phi_J \rangle
\label{eq:HIJBOt}
\end{eqnarray} 
A given term in the sum can be read as: the electron $v$ creates an electrostatic potential on the tessera $j$, that induces a polarization charge on the tessera $i$ that, in turn, creates a potential acting on the electron $u$. Note that $u$ may be equal to $v$ (i.e., an electron interacts with its own polarization charge).

The BO limit would be computationally advantageous, since it corresponds to an effective Hamiltonian for the molecule that has no state-specific dependence,\cite{Hynes92,Basilevsky1994} and thus all the excited states could be found in a single diagonalization shot. Excitation energies (correct at the second order in solute-solvent interaction) would then be calculated as:
 \begin{eqnarray}
\Delta E_{I0,BO}^{(2)}&=&\Delta E_{I0} + \frac{1}{2} \sum_{ij} \langle\Phi_I| \hat{V}_i Q_{0,ij}  \hat{V}_j| \Phi_I\rangle -
\frac{1}{2} \sum_{ij} \langle\Phi_0| \hat{V}_i Q_{0,ij}  \hat{V}_j| \Phi_0 \rangle
\label{eq:DE_BO}
\end{eqnarray} 

The extension to solvent of generic polarity is discussed in Appendix \ref{app:polar}. 

We would like to remark that the assumption leading to the BO limit is a rather crude one: it amounts to consider that even the excitations from the inner cores to the outer valence states have much lower energy than those of the solvent. This is clearly an untenable assumption, since solvents generally absorbs in the UV, i.e., with excitation energies that are higher than just a few of the solute excitations. This point will be further elaborated at the end of this section on the theoretical side, while we shall show in the numerical part that the \textcolor{black}{BO regime} leads to a large overestimation of the solvation energy. 

{\it BO Hamiltonian and its relation with solute quantum fluctuations}. We have recalled here above the interpretation of the BO Hamiltonian in terms of dielectric screening of the electron-electron interaction. Let us focus on the $H_{00}^{BO}$ matrix element, that represents a second order perturbative approximation to the solvation energy of the ground state:

\begin{eqnarray}
H_{00}^{BO}&=&E^{(2)}_{0,BO}=E_0 + \frac{1}{2} \sum_{ij} \langle\Phi_0| \hat{V}_i Q_{0,ij}  \hat{V}_j| \Phi_0 \rangle
\label{eq:H00BO}
\end{eqnarray}

Eq.(\ref{eq:H00BO}) provides the starting point for a rather transparent interpretation of the BO solvation energy. To make it physically intuitive, we simplify the PCM description to the Onsager one. As noted before,\cite{Cammi2005,Corni2005}  this amounts to replace $\hat{\mathbf{V}}$ with $\hat{\vec{\mu}}$ (the dipole moment operator) and $Q_{0,ij}$ with $g_0=-(\epsilon_0-1)/((2\epsilon_0+1)R^3)$, where $\epsilon_0$ is the dielectric constant of the solvent and $R$ the radius of the spherical cavity that in the Onsager model hosts the solute. With this simplifications we can write:
\begin{eqnarray}
E^{(2)}_{0,BO,Ons}&=&E_0 - \frac{1}{2} g_0 \langle\Phi_0| \hat{\vec{\mu}} \cdot \hat{\vec{\mu}} | \Phi_0 \rangle 
\label{eq:H00BO_Ons}
\end{eqnarray}
that, by adding and subtracting the term $\frac{1}{2}g_0 |\langle \Phi_0| \hat{\vec{\mu}}| \Phi_0 \rangle |^2$=$ \frac{1}{2}g_0 |\vec{\mu}_{00}|^2 $ led to:
\begin{eqnarray}
E^{(2)}_{0,BO,Ons}&=&E_0 -\frac{1}{2}g_0 |\vec{\mu}_{00}|^2 - \frac{1}{2} g_0 \langle \Phi_0| \left(\hat{\vec{\mu}} \cdot \hat{\vec{\mu}}-|\vec{\mu}_{00}|^2 \right) | \Phi_0 \rangle 
\label{eq:H00BO_Ons2}
\end{eqnarray}
The second term in the r.h.s. is the usual polarization contribution, the third one contains the quantum-mechanical fluctuation of the solute dipole. It clearly looks like a term that belong to what is generally identified as dispersion, rather than polarization. We shall show in sec. \ref{sec:disp} that the Hamiltonian term in eq.(\ref{eq:HIJ}) that leads to the BO form is indeed a term accounting for dispersion. 

{\it A refined form of the BO approximation.} 
To conclude this section, we describe a possible refinement to the BO-like approximation, where the assumption of fast solvent response is used only for a subset of the excitations.  In particular, R.A. Marcus proposed to separate the behavior of the transferring electron from that of the core electrons.\cite{Marcus1992} For the latter, the SCF approximation is most suitable since core electrons have excitation energies that are not small compared to solvent ones. A similar partition was also proposed in ref. \citenum{Basilevsky1994}. These kind of approximations turn out to be quite natural in the framework we have developed: they amount to consider $\mathbf{Q}'(\omega_{A'K})=\mathbf{Q}_0$ for a set of states $K$ whose energy is below a threshold $E_{thres}$ chosen on the basis of solvent excitation energies, and $\mathbf{Q}'(\omega_{A'K})=0$ for $K$-states with higher energy. More specifically, let us assume polyelectronic state $\Phi_K$ that are Slater determinants, obtained from a reference state $\Phi_0$ that can be the Hartree Fock determinant. Following ref. \citenum{Marcus1992}, we can distinguish the orbitals of the core $\phi_c$ from the ones describing the "slow electron" in $\Phi_0$, dubbed $\phi_h$ (h=HOMO). The "slowness" of the electron translates into the existence of several states  $\Phi_K=\Psi_h^r$ (obtained from $\Phi_0$ by single substitution of the occupied orbital $\phi_h$ with virtual orbitals $\phi_r$ with $r=LUMO,LUMO+1,...$) for which $\omega_{K0}$ is below $E_{thres}$. On the contrary, all the excitations from the core lead to excitation energies larger than  $E_{thres}$. Starting by eq.(\ref{eq:H00s}), the ground state energy becomes:
\begin{eqnarray}
\nonumber E^{(2)}_0&=&E_0 + \frac{1}{2} \sum_i q_i^{core}V_i^{core} + \frac{1}{2} \sum_i q_i^{hh}V_i^{hh} +\sum_i q_i^{core}V_i^{hh}+ \\
&+& \frac{1}{2} \sum_{r=LUMO,...}\sum_i q_i^{hr}V_i^{hr}
\label{eq:H00m}
\end{eqnarray}
where $\mathbf{V}^{core}$ and $\mathbf{q}^{core}$  are the electrostatic potential from the core electrons and the corresponding static apparent surface charges, $\mathbf{V}^{hh}=\langle\phi_h| \hat{\mathbf{V}} | \phi_h \rangle$ is the potential generated by the charge distribution corresponding to orbital $\phi_h$ (and $\mathbf{q}^{hh}$ is the corresponding apparent surface charge set) and $\mathbf{V}^{hr}=\langle\phi_h| \hat{\mathbf{V}} | \phi_r \rangle$,  $\mathbf{q}^{hr}=\mathbf{Q}_0 \mathbf{V}^{hr}$. If we now assume that the sum in $r$ could be extended to the core orbitals as well (i.e., that the additional contribution to the sum would be negligible), we can reformulate eq.(\ref{eq:H00m}) as:
\begin{eqnarray}
\nonumber E^{(2)}_0&=&E_0 + \frac{1}{2} \sum_i q_i^{core}V_i^{core} +\sum_i q_i^{core}V_i^{hh}+ \\
&+& \frac{1}{2} \sum_{ij} \langle\phi_h| \hat{V}_i Q_{0,ij}\hat{V}_j | \phi_h \rangle
\label{eq:H00m2}
\end{eqnarray}
where the last (BO-like) term is originated by the third and the last terms on the r.h.s of eq.(\ref{eq:H00m}). In eq.(\ref{eq:H00m}), the description of core electrons and of their interaction with the slow electron is SCF-like, while the slow electron is described by a BO-like term. This description has been suggested by Marcus in ref.\citenum{Marcus1992}, and we can also envisage to extend it from one to a few electron-hole transitions. For example, the natural transition orbital analysis may guide the choice of such transitions.\cite{Martin2003}   Notice however that to reach eq.(\ref{eq:H00m2}), \textcolor{black}{it is not enough to assume the existence of low-lying excited states that define the electron as slow. We have to include also the stronger hypothesis that} 
the other possible excitations of the electron (including the de-excitations to the core level) does not contribute significantly to the sum, allowing the use of a closure relation. Justifying such approximation is not trivial; certainly it should work better for excitation energies since these additional terms for the excited and the ground state should partially cancel out.
 
\subsection{Solute-solvent dispersion  \textcolor{black}{interaction: a formulation based on} real-frequency response functions}
\label{sec:disp}

So far, we have left open the problem of the physical meaning of some of the terms in previous equations, in particular eqs.(\ref{eq:H00s})-(\ref{eq:H00sij}). To address it, we shall preliminary work out an expression for the solute-solvent dispersion interactions where the response functions written for the real frequency $\omega$ (i.e., $\mathbf{Q}(\omega)$ for the solvent) appear. The resulting equation is not specific of OQS-PCM, but it is a general expression (new in the context of solute-solvent interaction studies) that can be used to rationalize and to calculate dispersion energies.

Considering the molecule (system) and the solvent (bath) as interacting quantum systems, the starting point of our derivation is the expression for dispersion interaction energy (in the ground state to simplify the notation) given by intermolecular interaction theory following McWeeny\cite{Mcweeny1992methods}: 
\begin{eqnarray}
E_{dis}=\frac{1}{2}\sum_{K\neq0,P\neq0}\sum_{ij} \frac{\langle \Phi_0 \Upsilon_0| \hat{V}_{i}\hat{q}_{i}| \Phi_K \Upsilon_P \rangle \langle \Phi_K \Upsilon_P | \hat{V}_{j}\hat{q}_{j}| \Phi_0 \Upsilon_0 \rangle  }{\Delta E_{K0} + \Delta E_{P0}}
\label{eq:dis}
\end{eqnarray}
In eq.(\ref{eq:dis}), we have explicitly written the interaction as in eq.(\ref{eq:int}) in terms of the operators $\hat{V}_{i}$ and $\hat{q}_{i}$, $K$ runs on the unperturbed solute states $| \Phi_K  \rangle$, $P$ on the unperturbed solvent state $|\Upsilon_P \rangle$, $\Delta E_{K0}$ is the solute excitation energy to the state $K$ and $\Delta E_{P0}$ is the solvent excitation energy to the state \textcolor{black}{$P$}. 
In Appendix \ref{sec:appA} we show the traditional approach of Quantum Chemistry to rearrange eq.(\ref{eq:dis}) by exploting a specific integral relation, leading to (see Appendix \ref{sec:appA} for the derivation):
\begin{eqnarray}
E_{dis}=\frac{1}{\pi}\int_0^{\infty}d\omega ~\sum_{K\neq0}\frac{\Delta E_{K0}}{\Delta E_{K0}^2+\omega^2}\int d\vec{r}_1 \sum_{ij} \frac{1}{|\vec{r}_{1}-\vec{s}_i|} \rho(K0|\vec{r}_1) Q_{ij}(i\omega)V_j(K0)
\label{eq:dis_int}
\end{eqnarray}
where $\rho(K0|\vec{r}_1)$ is the transition density of the solute for the $0-K$ excitation and $V_j(I0)$ is the corresponding potential of the tessera $j$. $Q_{ij}(i\omega)$ is the corresponding frequency dependent PCM matrix evaluated at the imaginary frequency $i\omega$. This is the relation initially put forwards in references \citenum{Amovilli1994,Amovilli1997}. 

An alternative transformation of the denominator of eq.(\ref{eq:dis}) is reported in the Quantum Optics textbook by Cohen-Tannoudji et al.\cite{Cohen04}, that can be used to obtained a different expression for $E_{dis}$ as documented in Appendix \ref{sec:appB}:
\begin{eqnarray}
E_{dis}&=&E_{dis}^{Sflu,Bres}+E_{dis}^{Sres,Bflu}\label{eq:dis_cohen_2}\\
E_{dis}^{Sflu,Bres}&=&\frac{1}{2}\int \frac{d\omega}{2\pi}\sum_{ij}  C_{mol,ij}^{sym} (\omega) Q'_{ij}(\omega)\\
E_{dis}^{Sres,Bflu}&=& \frac{1}{2}\int \frac{d\omega}{2\pi} \sum_{ij} \chi _{mol,ij}'(\omega)C_{ij}^{sym} (\omega)
\end{eqnarray}

where we have introduced $C_{mol,ij}^{sym} (\omega)$, the Fourier transform of the molecular symmetric correlation function of the electrostatic potential operators ($\hat{V}_i$ and $\hat{V}_j$), and the Fourier transform of the corresponding response function, $\chi_{mol,ij}(\omega)$. Their expressions are given by:
\begin{eqnarray}
\nonumber C_{mol,ij}^{sym} (\omega) &=& \Theta(-\omega)\chi_{mol,ij}''(\omega)-\Theta(\omega)\chi_{mol,ij}''(\omega)=\\ 
&=&\pi \sum_{K\neq0} \langle \Phi_0 | \hat{V}_{i}| \Phi_K \rangle \langle \Phi_K  | \hat{V}_{j}| \Phi_0 \rangle \left( \delta (\omega+\Delta E_{K0}) + \delta (\omega-\Delta E_{K0}) \right) \\
\chi_{mol,ij}'(\omega)&=&\left( \sum_{K\neq 0} \principalvalue \frac{\langle \Phi_0| \hat{V}_{i}| \Phi_K  \rangle \langle \Phi_K | \hat{V}_{j} | \Phi_0 \rangle  }{\Delta E_{K0} + \omega} + \principalvalue \frac{\langle \Phi_0| \hat{V}_{i}| \Phi_K  \rangle \langle \Phi_K | \hat{V}_{j} | \Phi_0 \rangle  }{\Delta E_{K0} - \omega}\right)
\end{eqnarray}

It should be noted that in the definition of $C_{mol,ij}^{sym} (\omega)$  we have left out the term $K=0$, which is responsible of the polarization, rather than the dispersion, interaction. Eq.(\ref{eq:dis_cohen_2}) yields a very intuitive picture of dispersion: the first term ($E_{dis}^{Sflu,Bres}$) accounts for the {\it res}ponse of the solvent ({\it B}ath) to the {\it S}olute (quantum) {\it flu}ctuations, the second term ($E_{dis}^{Sres,Bflu}$) accounts for the response of the solute to the solvent (quantum) fluctuations. This form, that involves integration over real frequency and separate fluctuations ($\rightarrow$ symmetric correlation functions) from response to an external perturbation ($\rightarrow$ response functions), is more intuitive than the relation eq.(\ref{eq:dis_int}). It is also more straightforwardly connected to the textbook description of dispersion as instantaneous dipole-induced dipole interactions, as this is indeed the leading multipolar contributions of the two terms in eq.(\ref{eq:dis_cohen_2}). The same expression, for generic system/reservoir pairs, was already presented in ref.\citenum{Cohen04}, \textcolor{black}{but here we clearly point out its relationship with dispersion interaction, specifically in the case of solute/solvent systems.} It is also noteworthy that an expression of $E_{dis}$ equivalent to the first term in the r.h.s. of eq.(\ref{eq:dis_cohen_2}) for dipolar solutes in spherical cavities has been proposed in the '60s by Linder.\cite{Linder1962} It was in fact based on physical considerations related to (quantum) fluctuations of molecular charge distribution and the solvent response induced by such fluctuations. Notably, the other portion of the dispersion energy ($E_{dis}^{Sres,Bflu}$ in eq.(\ref{eq:dis_cohen_2})) was missing as no fluctuation of the solvent reaction field was introduced.\\
\textcolor{black}{We finally note that in the OQS-PCM formalism here developed there are no Pauli repulsion effects, which tend to compensate the attractive dispersion interactions. The main reason is due to the assumption, in the theory of OQS, of non-overlapping wave functions of system and bath.
Indeed, the repulsion forces between two interacting molecules originates mainly from the Pauli exclusion principle, increasing
with the overlap of the two distributions and being related to the density of electrons with the same spin. 
A possible way to get around this limit could be to add \emph{a posteriori} a repulsion contribution,   for instance taking into account the results from the perturbation theory of molecular interactions adapted to account for electron antisymmetry, such as the symmetry adapted perturbation theory (SAPT) approach.\cite{Jeziorski} Another possibility is to introduce the supersystem (i.e. system+bath) wave function description in terms of an antisymmetrized group function into the open quantum system formalism, in a similar spirit of the work done for the densities of interacting fragments by McWeeny\cite{McWeeny59,Mcweeny63,Mcweeny1992methods} and Amovilli.\cite{Amovilli90} This last option will be considered as a development of the present work.} 
\subsubsection{Dispersion and State Specific PCM Hamiltonian.} 
\label{sec:dis_pcm}
Dispersion energy as written in eq.(\ref{eq:dis_cohen_2}) contains two terms. Comparing them with the diagonal QOS-PCM Hamiltonian element eq.(\ref{eq:H00s}), we find that $E_{dis}^{Sflu,Bres}$is indeed the same that we have isolated in the third line of the diagonal QOS-PCM Hamiltonian element. We therefore recognize that such term is due to solute-solvent dispersion interaction. A similar term with the same physical origin can also be recognized in the diagonal QOS-PCM Hamiltonian element of the excited states, eq.(\ref{eq:DE2}). However, neither excited nor ground state QOS-PCM energies contains the {\it second} of the two terms in the dispersion energy, $E_{dis}^{Sres,Bflu}$. As noted before, such term arises \textcolor{black}{from} the solute response to solvent fluctuations. Seen in this light, it is not surprising that our treatment missed such term: at the very beginning of \textcolor{black}{our time-independent OQS-PCM} derivation, we explicitly neglected the stochastic term accounting for solvent fluctuations (see text before eq.(\ref{eq:SSEs})), and here we find a consequence of that approximation. Of course, the solution of the full SSE would include also the missing term. Alternatively, if one wants to stick to a treatment based on stationary states, such missing term should be properly accounted for. Actually this is the standard PCM approach (i.e., adding \textcolor{black}{an} explicit dispersion term on top of polarization). 
Finally, there is also another consequence in the partial accounting of dispersion interaction yielded by the present QOS-PCM Hamiltonian. It was noted before that the correction to excitation energies found in PCM linear response theories is part of what has been described with dispersion.\cite{Corni2005}   The different way dispersion is written here does not make clear this point, and the linear response like term does not appear explicitly in the QOS-PCM excitation energies expression eq.(\ref{eq:DE2}). Yet, as we discussed in this paragraph, from the present form of dispersion one can also go back to the former treatment of dispersion, where the interplay between the linear response term and dispersion is already established. In fact, the linear response term is part of the solute response-solvent fluctuation term that is missing due to our initial assumption on the stochastic term, $E_{dis}^{Sres,Bflu}$. Taking this term and expanding the definition of $\chi _{mol,ij}'(\omega)$ and $C_{ij}^{sym} (\omega)$:
\begin{eqnarray}
\nonumber E_{dis}^{Sres,Bflu}&=&\frac{1}{2}\int \frac{d\omega}{2\pi} \sum_{ij} \chi _{mol,ij}'(\omega)C_{ij}^{sym} (\omega)=\\
\nonumber &&\frac{1}{2}\int \frac{d\omega}{2\pi} \sum_{ij} \left( \sum_{K\neq 0} \principalvalue \frac{\langle 0| \hat{V}_{i}| \Phi_K  \rangle \langle \Phi_K | \hat{V}_{j} | 0 \rangle  }{\Delta E_{K0} + \omega} + \principalvalue \frac{\langle 0| \hat{V}_{i}| \Phi_K  \rangle \langle \Phi_K | \hat{V}_{j} | 0 \rangle  }{\Delta E_{K0} - \omega}\right) \cdot \\
&&\cdot \pi \sum_{P\neq0} \langle \Upsilon_0 | \hat{q}_{i}| \Upsilon_P \rangle \langle \Upsilon_P  | \hat{q}_{j}| \Upsilon_0 \rangle \left( \delta (\omega+\Delta E_{P0}) + \delta (\omega-\Delta E_{P0}) \right) 
\label{eq:sres}
\end{eqnarray}

which can be simplified (for real wavefunctions) as:
\begin{eqnarray}
\nonumber E_{dis}^{Sres,Bflu}&=&\frac{1}{2} \sum_{K\neq 0,P\neq0,ij} \left(  \frac{\langle 0| \hat{V}_{i}| \Phi_K  \rangle \langle \Phi_K | \hat{V}_{j} | 0 \rangle  }{\Delta E_{K0} + \Delta E_{P0}} + \frac{\langle \Phi_0| \hat{V}_{i}| \Phi_K  \rangle \langle \Phi_K | \hat{V}_{j} | \Phi_0 \rangle  }{\Delta E_{K0} - \Delta E_{P0}}   \right)
\langle \Upsilon_0 | \hat{q}_{i}| \Upsilon_P \rangle \langle \Upsilon_P  | \hat{q}_{j}| \Upsilon_0 \rangle=\\
\nonumber  &&\frac{1}{2} \sum_{K\neq 0,P\neq0,ij} \left(  \frac{ \langle \Upsilon_0 | \hat{q}_{i}| \Upsilon_P \rangle \langle \Upsilon_P  | \hat{q}_{j}| \Upsilon_0 \rangle }{\Delta E_{K0} + \Delta E_{P0}} + \frac{\langle \Upsilon_0 | \hat{q}_{i}| \Upsilon_P \rangle \langle \Upsilon_P  | \hat{q}_{j}| \Upsilon_0 \rangle  }{\Delta E_{K0} - \Delta E_{P0}}   \right) \langle \Phi_0| \hat{V}_{i}| \Phi_K  \rangle \langle \Phi_K | \hat{V}_{j} | \Phi_0 \rangle \\
 \label{eq:gs_lr}
 \end{eqnarray}
The same treatment can be applied to the solute excited state $I$, leading to:
 \begin{eqnarray}
 \nonumber E_{dis, I}^{Sres,Bflu}&=&\frac{1}{2} \sum_{ij} \sum_{K'\neq I,P\neq0} \left(  \frac{ \langle \Upsilon_0 | \hat{q}_{i}| \Upsilon_P \rangle \langle \Upsilon_P  | \hat{q}_{j}| \Upsilon_0 \rangle }{\Delta E_{K'I} + \Delta E_{P0}} + \frac{\langle \Upsilon_0 | \hat{q}_{i}| \Upsilon_P \rangle \langle \Upsilon_P  | \hat{q}_{j}| \Upsilon_0 \rangle  }{\Delta E_{K'I} - \Delta E_{P0}}   \right) \langle \Phi_I| \hat{V}_{i}| \Phi_K'  \rangle \langle \Phi_K' | \hat{V}_{j} | \Phi_I \rangle \\
 \label{eq:es_lr}
 \end{eqnarray}
 Let us now extract from eqs.(\ref{eq:gs_lr}) and (\ref{eq:es_lr}) the terms that contains the $0-I$ transition density, and let us take their difference to find the linear response like contribution to the excitation energy $\Delta E_{LR}$. This amounts to extract the term with $K=I$ from eq.(\ref{eq:gs_lr}) and the term with $K'=0$ from eq.(\ref{eq:es_lr}). Their difference reads:
  \begin{eqnarray}
  \nonumber \Delta E_{LR}= \sum_{ij} \sum_{P\neq0} \left(  \frac{ \langle \Upsilon_0 | \hat{q}_{i}| \Upsilon_P \rangle \langle \Upsilon_P  | \hat{q}_{j}| \Upsilon_0 \rangle }{\Delta E_{P0} - \Delta E_{I0}} - \frac{\langle \Upsilon_0 | \hat{q}_{i}| \Upsilon_P \rangle \langle \Upsilon_P  | \hat{q}_{j}| \Upsilon_0 \rangle  }{\Delta E_{I0} + \Delta E_{P0}}   \right) \langle \Phi_I| \hat{V}_{i}| \Phi_0  \rangle \langle \Phi_0 | \hat{V}_{j} | \Phi_I \rangle \\
\end{eqnarray}

When (some of the) solvent excitation energies are in the same range as solute ones, the first term in the sum dominates and we can approximate $\Delta E_{LR}$ as:
\begin{eqnarray}
  \Delta E_{LR}= \sum_{ij} Q_{ij}(\omega_{0I}) \langle \Phi_I| \hat{V}_{i}| \Phi_0  \rangle \langle \Phi_0 | \hat{V}_{j} | \Phi_I \rangle
\end{eqnarray}
which is exactly the linear response term found before.\cite{Corni2005} On the contrary, when all the solvent energies are much larger than $\Delta E_{I0}$, the two terms in the sum cancel out each other. This is fully in line with the interpretation of the linear response term as a excitonic energy:\cite{Lunkenheimer} when the solvent excitations are far high, they do not provide differential stabilization to the ground vs the excited states of the solute. 

\subsubsection{Dispersion and solvent BO Hamiltonian.} 
Let us now analyze eq.(\ref{eq:dis_cohen_2}) to show its relation with the solvent BO limit. $E_{dis}^{Sflu,Bres}$ is the part of the sum that, for the ground state, provides the BO Hamiltonian (eq.(\ref{eq:HIJBO})) when we take the limit $Q'_{ij}(\Delta E_{K0})\rightarrow Q'_{ij}(0)$ for any solute excitation $0-K$ in eq.(\ref{eq:HIJ}).  $E_{dis}^{Sres,Bflu}$ in eq.(\ref{eq:sres}) is negligible in the same limit: $C_{ij}^{sym} (\omega) $ is different from zero only for frequency $\omega$ close to the excitation energies of the solvent (see fluctuation-dissipation theorem in eq.(\ref{eq:fd})). Within the assumption that such energies are much larger than any excitation energy of the solvent, the real part of the solvent response function goes to zero in this frequency range (physically: the solute cannot follow the fluctuations of the solvent) and the integrand in eq.(\ref{eq:sres}) (and thus $E_{dis}^{Sres,Bflu}$) is therefore zero. In summary, we find that the solvent Born-Oppenheimer limit is incorporating the dispersion energy term as well, although in a specific (and strongly approximated) limit. In the numerical section we shall provide an estimate of the quality of such approximation.

\section{Numerical Results}
\label{sec:numerical}
\begin{figure}[h!]
    \centering
    \includegraphics[width=0.7\textwidth]{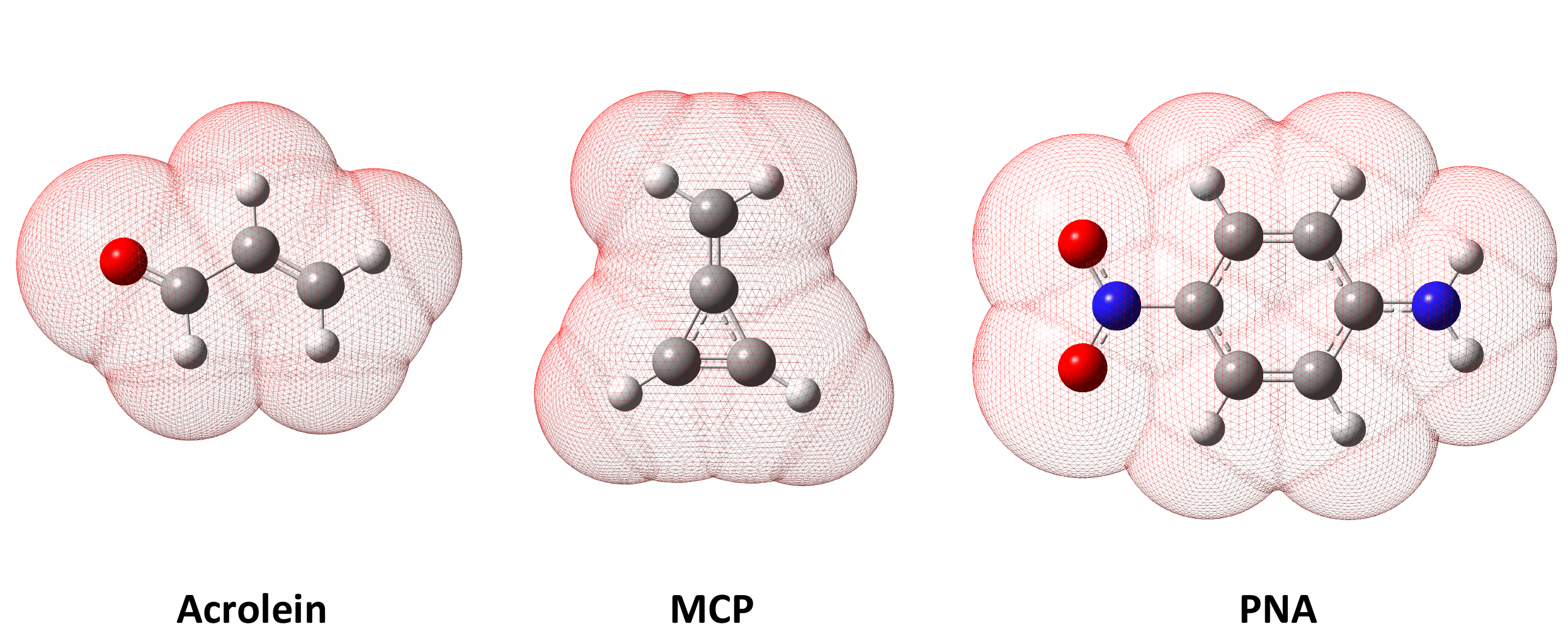}
    \caption{Structures of studied systems: acrolein, methylencyclopropene (MCP) and para-nitroaniline (PNA). The PCM cavity boundary is also shown: each tessera hosts an apparent surface charge $q_i$ as described in the main text.}
    \label{fig:systems}
\end{figure}
In this section we provide illustrative numerical results of the key theoretical  developments presented  in this work. \textcolor{black}{In particular, in sect. \ref{sec:OQSsimul} we present solvation energy and solvatochromic shift results for an approximated version of time-independent OQS-PCM, based on eq.(\ref{eq:TI}) in the form proper for solvents of generic polarity (Appendix A). We also report there the results of standard PCM calculations, including dispersion. In sect. \ref{sec:BOsimul} we discuss instead BO solvation energies.}\\ 
\textcolor{black}{\emph{Solutes and solvents considered.} We simulate the ground and first two singlet excited states of 3 systems: acrolein, methylencyclopropene (MCP) and para-nitroaniline (PNA), whose structures are shown in fig.\ref{fig:systems}.
Two solvents with low (dioxane, $ \varepsilon=$ 2.21) and high polarity (acetonitrile, $ \varepsilon=$ 35.69) were used. GS geometries were optimized at DFT level, using the B3LYP \cite{B3LYP} functional and the 6-31+G(d) basis set, exploiting the G16 version \cite{g16} of GAUSSIAN suite of programs.}\\
\textcolor{black}{T}he calculation  of the elements of the OQS-PCM Hamiltonian in eq.(\ref{eq:HIJ}) would require the  evaluation of the frequency dependent solvent response  matrices $\mathbf{Q}'(\omega_{IK})$ and $\mathbf{Q}'(\omega_{0K})$  for all the pairs of states $I-K$ of the solute, \textcolor{black}{which is computationally demanding. Therefore,} we revert to an approximation\textcolor{black}{: we} assume that $\mathbf{Q}'(\omega)=\mathbf{Q}'_0$ (or $\mathbf{Q}'_d$ for polar solvents as detailed in Appendix \ref{app:polar}) if $\omega \leq \omega_{cut-off}$, and  $\mathbf{Q}'(\omega)=0$ if $\omega > \omega_{cut-off}$. 
Under this approximation, dubbed OQS$_a$ in the following, the 
OQS-PCM solvation energy is obtained by diagonalizing the matrix :
\begin{eqnarray}\label{eq:OQSa}
 H_{\Phi'_A,IJ}^{OQS}&\approx&E_I\delta_{IJ} +\frac{1}{2} \sum_{K=0}^{K_{cut-off}} \sum_{ij}Q'_{d,ij} \langle\Phi_I^f| \hat{V}_i |\Phi_K^f\rangle\langle \Phi_K^f | \hat{V}_j| \Phi_J^f \rangle.
\end{eqnarray}
\textcolor{black}{This expression corresponds to that in eq.(\ref{eq:HIJ}) for a solvent of generic polarity (see Appendix \ref{app:polar}), that  replaces  ${Q}'_{0,ij}$ with ${Q}'_{d,ij}$ and makes use of GS frozen reaction field states, $|\Phi_K^f\rangle$. The latter encompass gas-phase states as a special case. Indeed, these frozen states are better starting points than gas-phase ones for excited states and property calculations, because the polarization of the ground state orbitals is often one of the largest solvent-induced effects.}
To easily implement and analyze the approximated schemes adopted,  a CIS ansatz of the wave-functions have been used, coupled to the cc-pVTZ basis set.
 We remark that the goal of this analysis is the comparison between different solvation \textcolor{black}{energies} and thus even medium or low level calculations (such as CIS) can be sufficient. 
 All the CIS calculations have been carried out exploiting a locally modified  version of GAMESS-US code\cite{GAMESS1,GAMESS2}, whereas the cavities have been obtained exploiting the homemade TDPlas code, developed in our group.
All these data are the input of a python script that diagonalize \textcolor{black}{ eq.(\ref{eq:OQSa}), to obtain the entire spectrum of the approximate version of OQS energies.}
 The cut-off frequency ($\omega_{cut-off}$) is estimated as the first excitation energy of each solvent which correspond to the first CIS singlet of each solvent molecule, in a PCM environment. In this manner we obtain a consistent description of both solute and solvent characteristic frequencies.\\
\textcolor{black}{In the case of the BO calculations, the resolution of identity in the CIS-ansatz form reduces
 eq.(\ref{eq:HIJBO}) (or rather its equivalent from Appendix \ref{app:polar}) to:
 \begin{eqnarray}\label{eq:BOa}
	 H_{\Phi'_A,IJ}^{BO}&\approx&E_I\delta_{IJ} 
	+ \frac{1}{2} \sum_{K=0}^{CIS} \sum_{ij}Q'_{d,ij} \langle\Phi_I^f| \hat{V}_i |\Phi_K^f\rangle\langle \Phi_K^f | \hat{V}_j| \Phi_J^f \rangle
\end{eqnarray}
where the sum in K runs on all the CIS states. }
\textcolor{black}{We also tested the convergence of such sum over K, see sect. \ref{sec:BOsimul}.}
\\  
\textcolor{black}{ \emph{Details of conventional PCM calculations.} We simulate all the GS data employing the integral equation formalism of PCM (IEF-PCM).\cite{Cances1997} 
Since 
OQS-PCM includes one part of the dispersion interactions between the QM sub-system and the polarizable bath, as proved in sec.\ref{sec:disp}, we have also explicitly added such contribution to the standard PCM results by one of the approaches put forward in the past for molecules in implicit solvents: to be able to treat both ground state and excited states solvation, in the present work we have included it with the SMSSP model.\cite{Marenich2013} 
For the sake of comparison, the Floris-Tomasi model\cite{Floris89} for the GS case have been also taken into account.
Excited state (ES) energies have been calculated by adopting a state specific correction to the PCM linear response (cLR)\cite{caricato2006formation} excitation energies. In cLR, the ES density-dependent term (obtained in the presence of the frozen reaction field) is computed and added as a correction to the frozen-solvent excitation energy. An overview of the state of the art of the different state-specific methods implemented can be found in a recent perspective.\cite{guido2019}}


\subsection{ \textcolor{black}{OQS$_a$-PCM results}}
\label{sec:OQSsimul}
\begin{table}[!h]
    \centering
    \begin{tabular}{l|ccccc|r}
    \hline
    \hline
         \multicolumn{7}{c}{GS}  \\
         \hline
     Solvent & System & PCM & PCM+FT & PCM+SMSSP & OQS$_a$ & BO$_a$\\
      \multirow{3}{*}{ACN} & ACR & -0.22	&-0.47 &-0.34	& -0.32 & -2.94 \\
      & MCP &  -0.15 & -0.39	& -0.29	&	-0.18	&	-3.19 \\
      & PNA & -0.40	&  -0.82    & -0.71	&	-0.62	&	-1.64 \\
      \hline
      \multirow{3}{*}{1,4-Diox} & ACR & -0.11 & -0.44 & -0.23 & -0.15 & -3.30 \\
      & MCP & -0.06	&-0.40  &-0.21 	&	-0.27	&	-3.61 \\
      & PNA & -0.21	& -0.78 & -0.51	&	-0.60	&	-1.64 \\
      \hline
    \end{tabular}
    \caption{CIS/cc-pVTZ ground state solvation energies (eV) of Acrolein, MCP and PNA in acetonitrile and 1,4-dioxane. Reported values:  electrostatic IEF-PCM results ("PCM"), same values including  dispersion effects simulated by the Floris-Tomasi model (PCM-FT) and the SMSSP one ("PCM+SMSSP"), the approximated OQS results ("OQS$_a$"). For sake of comparison, the approximated Born-Oppenheimer results ("BO$_a$") are also shown.}
    \label{tab:GSsolv}
\end{table}

\begin{table}[!h]
    \centering
    \begin{tabular}{l|ccccc|r}
    \hline
    \hline
         \multicolumn{7}{c}{S$_1$ $\leftarrow$ S$_0$}  \\
         \hline
     Solvent & System & $\omega_0$ & cLR & cLR+SMSSP & OQS$_a$ & BO$_a$\\
      \multirow{3}{*}{ACN} & ACR & 0.03	& 0.00	& -0.14	& 0.03 & -0.32  \\
      & MCP & 0.19	&	0.05	&	-0.11	&	0.06	&	-0.47   \\
      & PNA & 	-0.39	&	-0.42	&	-0.78	&	-0.68	&	-0.91 \\
      \hline
      \multirow{3}{*}{1,4-Diox} & ACR & 0.02 & -0.01 & -0.15 & 0.03 & -0.39 \\
      & MCP & 0.11	&	0.00	&	-0.16	&	-0.11	&	-0.67 \\
      & PNA & 0.03	&	-0.02	&	-0.39	&	-0.30	&	-0.55 \\
      \hline
           \multicolumn{7}{c}{S$_2$ $\leftarrow$ S$_0$}  \\
         \hline
     Solvent & System & $\omega_0$ & cLR & cLR+SMSSP & OQS$_a$ & BO$_a$\\
      \multirow{3}{*}{ACN} & ACR &	-0.26	&	-0.26	&	-0.40	&	-0.40 & -0.85  \\
      & MCP &  -0.14	&	-0.21	&	-0.37	&	-0.30	&	-1.26 \\
      & PNA & 	-0.45	&	-0.43	&	-0.84	&	-0.50	&	-0.61 \\
      \hline
      \multirow{3}{*}{1,4-Diox} & ACR & -0.10	&	-0.09	&	-0.24	&	-0.15	&-0.78 \\
      & MCP & -0.04	&	-0.09	&	-1.11	&	-0.58	&	-1.34 \\
      & PNA & -0.25	&	-0.21	&	-0.63	&	-0.36	&	-0.48\\
      \hline
      \hline
    \end{tabular}
    \caption{CIS/cc-pVTZ excited state solvation energies (eV) of Acrolein, MCP and PNA in acetonitrile and 1,4-dioxane. Reported values:  solvent polarization frozen to the ground state equilibrated one ("$\omega_0$"), the corrected Linear Response  data ("cLR"), the same with the SMSSP dispersion added ("cLR+SMSSP") and the approximated OQS ("OQS$_a$") results. For sake of comparison, the approximated Born-Oppenheimer results ("BO$_a$") are also shown.}
    \label{tab:ESsolv}
\end{table}

\begin{figure}[h!]
    \centering
    \includegraphics[width=\textwidth]{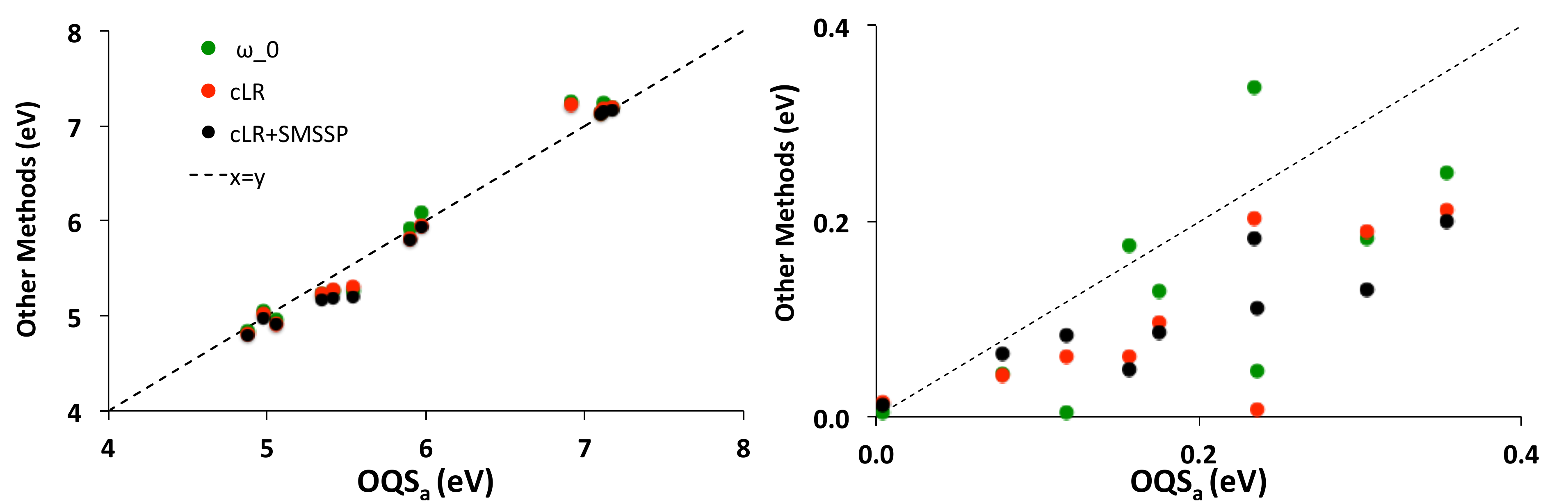}
    \caption{Correlation plots of other methods versus OQS$_a$ results (data in eV). Left: excitation energies. Linear regression R$^2$ values: $\omega_0=0.982$, cLR$=0.986$, cLR+SMSSP$=0.989$.
    Right:  gas to solvent excited state shifts. Linear regression R$^2$ values: $\omega_0=0.48$, cLR$=0.59$, cLR+SMSSP$=0.77$.
    All the data from Table \ref{tab:ESsolv} have been used to calculate the shifts, with the exception of MCP in dioxane, S2, for which the SMSSP contribution is anomalously large.}
    \label{fig:correlation}
\end{figure}

\textcolor{black}{We start our analysis with the computation of solvation energies, i.e. the amount of energy associated with dissolving a solute in a solvent, with a OQSa description.} GS results are shown in table \ref{tab:GSsolv}: in line with the analysis carried forward in the theoretical sections, OQS$_a$ values are in general closer to IEFPCM simulations that include also dispersion effects (that of SMSSP model in particular).
\textcolor{black}{This is also observed for the solvation energies of the excited states (reported in their approximated form in Table \ref{tab:ESsolv}) when compared with those obtained by the ground state frozen-solvent approximation ("$\omega_0$") and the state specific cLR approach, plus SMSSP dispersion ("cLR+SMSSP").}
The OQS$_a$ solvation energies are generally closer to the cLR+SMSSP results than to cLR or $\omega_0$ ones, although this behavior is less defined than for the GS results discussed above. We remark, however, that the solvation energy data are bounded between the pure electrostatic and those including the dispersion effects in all cases.
Finally, to explore the point in more detail, in Fig.\ref{fig:correlation} we present correlation plots of excitation energies (left panel) and solvatochomic shifts (right panel) for OQS$_a$ vs $\omega_0$, cLR and cLR+SMSSP. The maximum correlation coefficient, R, is indeed obtained for cLR+SMSSP calculations, being 0.989 for excitation energies and 0.77 for solvatochromic shifts.

\subsection{\textcolor{black}{BO-PCM results}}\label{sec:BOsimul}
\begin{figure}[h!]
    \centering
    \includegraphics[width=\textwidth]{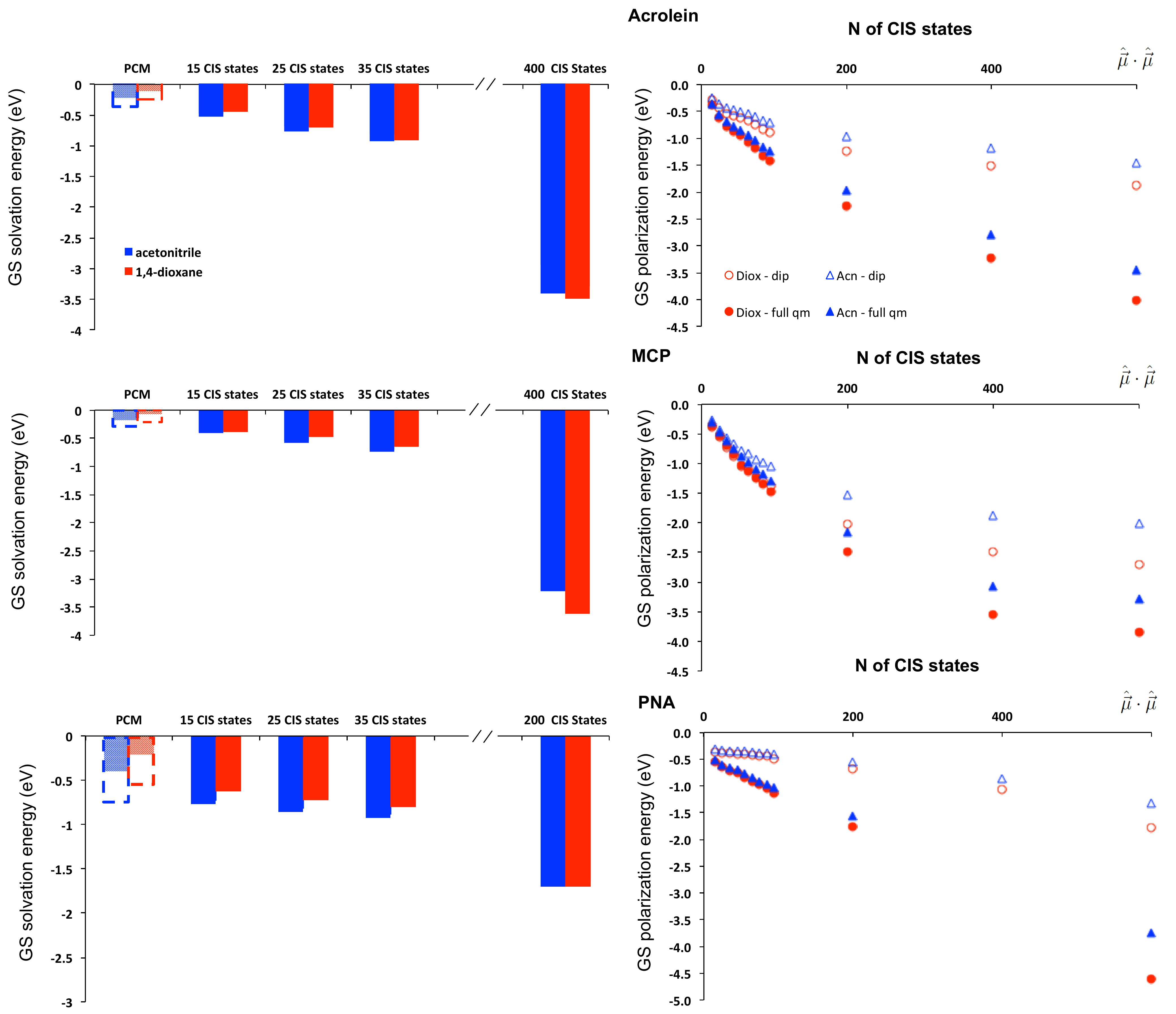}
    \caption{Left: CIS/cc-pVTZ approximated BO-PCM GS solvation energy (eV)  of acrolein (up), MCP (middle) and PNA (bottom) as a function of the number of states included in the calculation. In blue, values in acetonitrile; in red, values in 1,4-dioxane. Standard IEF-PCM calculations are also reported for comparison. Dashed lines includes also dispersion effects, evaluated by the SMSSP model\cite{Marenich2013}.  Right: approximated BO GS solvation energies (eV) for the Onsager model are compared to the approximated BO-PCM results. The {\it{limit}} BO values of the dipolar model are also reported ("$\vec{\mu}\cdot\vec{\mu}$"). The corresponding BO-PCM ones have been extrapolated.}
    \label{fig:GSsolv}
\end{figure}


Let us now analyze solvation energy results in the BO regime.
Preliminarily, in Fig.\ref{fig:GSsolv} we have explored how the GS solvation energies for the different systems change with the number of CIS excitations included in the sum over $K$ of eq.(\ref{eq:BOa}). It is apparent that approximated BO solvation energies are larger than the corresponding PCM+SMSSP values already for K $>15$ CIS states. 
Focusing on the convergence of the energy, from the figure \ref{fig:GSsolv}, in particular the right panels, it seems that such convergence is very slow, showing that to calculate the BO limit the sum over states approach is not effective. To better investigate this behaviour, we 
compare the full PCM results with those obtained considering a simpler case, the Onsager model used to derive the eq.(\ref{eq:H00BO_Ons2}), 
as reported in right panels of Fig. \ref{fig:GSsolv}. The Onsager approximated BO solvation energy (BO-Onsager) is qualitatively in agreement with the corresponding BO-PCM one in all the cases, especially for the lowest number of excitations included. There is however a quantitative discrepancy that increases by increasing the number of CIS electronic excitations included. This is in line with expectations: the higher the excited state, the more diffuse and/or featured will be the corresponding ground to excited transition density, and the poorer will be the dipole approximation. The use of the Onsager model easily allows the calculation of the exact limit of the resolution of identity sum: indeed, the expectation values of the operator $\hat{\vec{\mu}}\cdot \hat{\vec{\mu}}=\hat{x}\hat{x}+\hat{y}\hat{y}+\hat{z}\hat{z}$  that enters in the BO-Onsager limit, in eq.(\ref{eq:H00BO_Ons}), can be obtained as the sum of the diagonals element of the expectation values of the quadrupole operator that is routinely printed by Quantum Chemistry codes such as Gaussian. Using such data, we get as the Onsager model asymptotic values  1.9 eV and 1.5 eV for acrolein in 1,4-dioxane and acetonitrile, respectively. The analogous quantities obtained with the resolution of identity method using 400 states reached 1.5 eV and 1.2 eV, showing that indeed the sum converges slowly (20\% of difference), if it converges at all. We have in fact to take into account that we are using a finite basis set, and therefore there is no reason to converge exactly. The difference is even larger in the case of PNA (around 40\% in 1,4-dioxane, 35\% in acetonitrile). 

Incidentally, we note that the BO limit largely overestimate the dispersion corrected PCM: as reported in the left panel of fig.\ref{fig:GSsolv} and in Table \ref{tab:GSsolv}, the values of GS solvation energy in the BO limits are around 7 times larger than the PCM results for Acrolein, 8 times for MCP and 4 times for PNA. Such values (ranging from -1.6 eV to -3.6 ev, i.e., -160 kJ/mol to 360 kJ/mol) are unphysically large. While the BO limit would be very convenient as we could have a method that does not require state specific solutions of the electronic structure problem, and that takes into account dispersion in the determination of electronic wave-functions, these results are rather discouraging. 
A hint of such overestimation might also be seen in the GW results reported in ref.\onlinecite{Duchemin2016}, that slightly but consistently overestimate the solvation effect on ionization potential with respect to standard PCM. Of course, the overestimation is expected to be small, since in taking the difference between the neutral and the ionized form of the solute the large BO terms largely cancel out.  

 \textcolor{black}{Passing now to BO solvation energies of excited states (tab. II), they are also} very large, in an unphysical way, compared to all the PCM results.  Yet, the BO excited states solvation energies are consistently much smaller \textcolor{black}{than} for the ground state. We have identified this as an artifact of CIS applied to the sum over state BO results: in fact, the excited states entering such sum for the ground state are all the single excitations from the ground state itself. On the contrary, when calculating the BO term for a CIS excited state, \textcolor{black}{there is no single excitation} from such state in the sum (as they would be double excitations from the ground state, and thus not accounted for in the CIS methods). The BO sum over state is therefore a worse approximation for the excited states than for the ground states. We cannot therefore derive a conclusion  on the quality of the BO description for solvatochromic shifts, not even 
 \textcolor{black}{for} few systems studied here. While the accuracy of BO is certainly unacceptable to calculate absolute solvation energy, the accuracy of a mixed method using standard PCM for the ground state and BO to calculate solvatochromic shifts remains a numerically open question to be assessed on a significant database of molecules, ideally by skipping the resolution of identity approximation.

We point out that the unbalanced treatment of ground and excited states of CIS is much less relevant for the OQS$_a$ results reported in the previous section, at least for the lowest excited states analyzed here: in the excitation energy window considered by the present cut-offs, most of the excited states will be other CIS states from the ground states, rather than states with double excitations. The agreement between OQS$_a$ and cLR+D is therefore robust with respect to this CIS limitation.

\begin{figure}
    \centering
    \includegraphics[width=\textwidth]{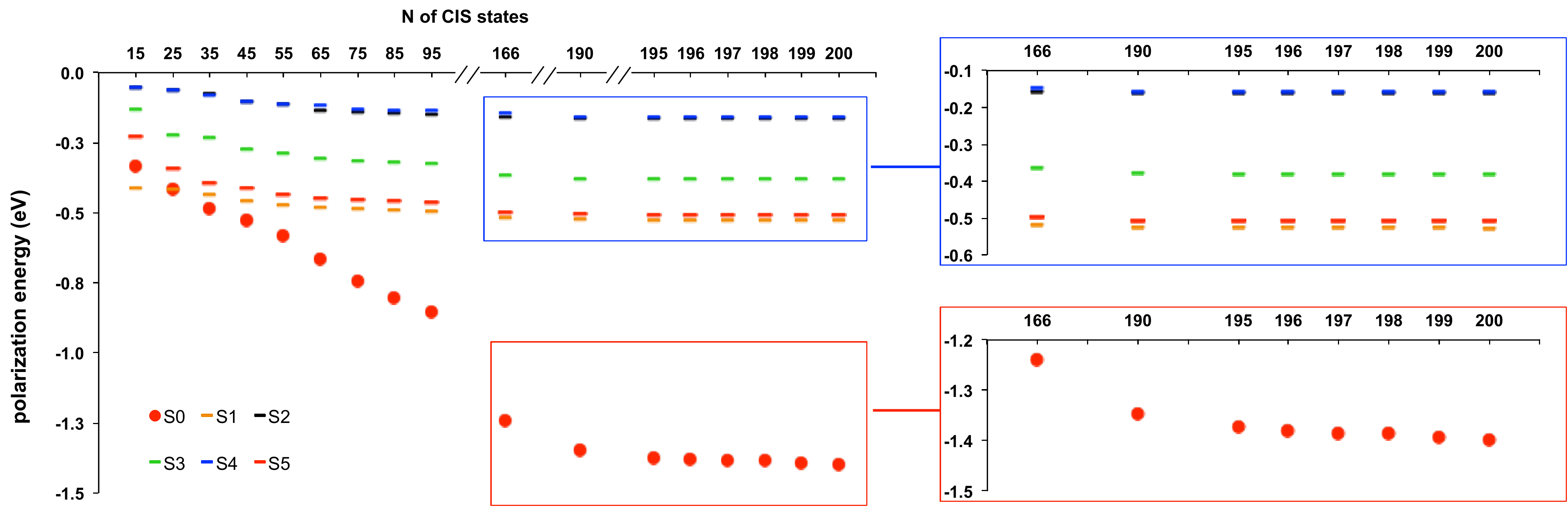}
    \caption{PNA in acetonitrile: convergence of the approximated BO-PCM solvation energy contribution to the BO limit as a function of the number of CIS states included. Color code: orange, S1 state; black, S2 state;
    green, S3 state;
    blue, S4 state;
    red, S5 state; red dots, ground state. }
    \label{fig:convergence}
\end{figure}

To show numerically how the differential BO solvation energy for the ground and excited states set in as a function of the number of states considered in the sum over states, we refer to fig.\ref{fig:convergence}, which shows the approximated BO-PCM solvation energy for the ground ("S0") and the first five excited states upon increasing the number of CIS states in the case of PNA in acetonitrile. The different trend of excited and ground state is evident, and it is also the reason why the first five singlets polarization energies converge to the BO limit (differences around $10^{-2}$ eV) already at 166 CIS states, whereas the GS energy needs at least 198 states. 

\section{Conclusions}
\label{sec:conclusions}
In this work we have presented a derivation of a PCM description of a quantum solute in a polarizable dielectric bath starting from the theory of OQS, leading to a OQS-PCM time-dependent Schr\"odinger equation that naturally accounts for the time scales of both solute and solvent coupled dynamics. We have then focused on stationary (ground and excited) states of the solute, and derived that they solve a time-independent Schr\"odinger equation where an effective Hamiltonian (here called OQS-PCM Hamiltonian) appears. Based on such Hamiltonian, we have obtained expressions of solvation energies for the ground and the excited states, finding that they confirm previous PCM results, and the related discussion on what terms the excitation energies in solution should contain (the PCM-SS terms, dispersion, and the excitonic-like PCM-LR term; the latter may be contained in the dispersion term or not depending on the expression used for dispersion).\cite{Cammi2005,Corni2005,Lunkenheimer,Duchemin2018} This is remarkable as OQS-PCM accounts, by construction, for the proper coupled time scales of solute and solvent electronic dynamics. In particular, we have recognized that one of the terms appearing in the OQS-PCM Hamiltonian is accounting for solute-solvent dispersion interaction. This led us to write an expression for such interaction based on mutual solute charge density fluctuations and consequent induced polarizations in the solvent (and viceversa) that is particularly close to the General Chemistry textbook intuitive definition of dispersion interaction.  Such expression also revealed that neglecting stochastic terms in the time-dependent OQS-PCM Schr\"odinger equation amounts to also neglect part of the solute-solvent dispersion interaction. \textcolor{black}{The OQS-PCM formalism here developed does not include terms connected to the Pauli repulsion effects, which often tend to reduce the attractive dispersion interactions, due to the non-overlapping assumption on wave function partition in the theory of open quantum systems. It is an important point for further developments.} 

While OQS-PCM can in principle be implemented with any quantum mechanical level of calculation for the solute, we have provided numerical examples of the OQS-PCM results, limited to CIS level, 
solvation energies of ground and excited states are in line with the theoretical analysis, showing that OQS-PCM provides results comparable to implicit solvent methods that account for polarization and dispersion interaction. 
Further, extensive, numerical investigation are required to evaluate the possible use of OQS-PCM as a practical method in Quantum Chemistry, besides being a formidably informative theoretical framework. \textcolor{black}{To this end, we are working on more efficient numerical implementation of the OQS-PCM method.} 
We anticipate that the extension of the OQS approach to QM/MM with polarizable force fields\cite{Mennucci2019} would also follow a similar scheme at provided here for PCM.\\
We have shown that the OQS-PCM includes (as a special case) the Born-Oppenheimer solvation regime\cite{Gehlen1992,Marcus1992,Hynes92,Basilevsky1994}, and that even refined versions of such regime suggested in the past\cite{Marcus1992} can be obtained from OQS-PCM by means of suitable approximations. As a corollary to that, we argued that the BO limit is therefore accounting not only for polarization but also for dispersion effect, although in a strongly approximated fashion. The BO limit has the intriguing property of lacking a state specific Hamiltonian, that would greatly simplified the PCM description of excited states. The preliminary numerical tests we performed highlighted the poor accuracy and unphysical results of this approach for absolute solvation energies of ground and excited states. However, we could not gauge the accuracy of the BO for solvatochromic shifts - where the large errors on the ground and the excited solvation energies might cancel out - due to the approximated approach (CIS+sum over state) used here (we remark that OQS-PCM is compatible with much more accurate theories, such as coupled cluster, as standard PCM)\cite{Cammi2009,Caricato2011}. Our theoretical analysis showed that the recently proposed GW/BSE PCM approach\cite{Duchemin2016,Duchemin2018} aimed at calculating solvatochromic shifts contains a (refined) BO physics; taking into account that such method delivered preliminary results of good accuracy,\cite{Duchemin2018} a method using standard PCM for the ground state and a BO approach to find energies in solution is also worthy to be developed and tested. Intermediate methods, that take into account the different time scales of solute electrons, are also an option to explore.\cite{Marcus1992}

In this work, we have not really exploited the full potential of the time-dependent OQS-PCM equations (involving stochastic terms as well). In fact, they provide a rigorous starting point for simulating not only the dynamics of molecules in solution (allowing, for example, a definition of excitation energies that closely matches the corresponding experimental counterpart), but also for molecules close to plasmonic nanostructures, the intrinsic dynamical timescale of whom is necessarily comparable to those of the molecule. However, they require a practical way to treat the resulting Non-Markovian SSE, which will be the topic of future investigations.

\appendix
\section{Extension to \textcolor{black}{solvent of generic polarity}}
\label{app:polar}
In the case of solvent of generic polarity, i.e. not limiting to apolar solvent,  we have to consider that the solvent (i.e., the environment) is initially equilibrated with a reference state for the solute (i.e., the system). Here we shall consider the solute ground state to be such a reference state. In this case, following Gaspard and Nagaoka \cite{Gaspard1999}, we can write the system Hamiltonian as:
\begin{eqnarray}
\hat{H}_s^f=\hat{H}_s+\sum_i q_i^0 \hat{V}_i
\end{eqnarray}
where $\hat{H}_s^f$ can be recognized as the PCM Hamiltonian with apparent charges frozen to the values proper for the ground state ($q_i^0$). We shall indicate the eigenstates of such Hamiltonian as $|\Phi_I^f\rangle$. In particular, $|\Phi_0^f\rangle$ is the usual (i.e., fully self-consistent) PCM solute ground state. 

In turn, still following Gaspard and Nagaoka, the operator $\hat{V}_i$ in the SSE must be replaced by $\hat{V}_i-V^0_i$ where $V^0_i$ is the expectation value of the electrostatic potential on the tessera i for the solute ground state. Performing such replacements and repeating the same derivations we presented for the apolar solvent case, we end up with analogous equations for the generic solvent case. Here we shall give only the results for excitation energy expressions recovered by the diagonal matrix element of the OQS-PCM Hamiltonian (i.e., eq.(\ref{eq:DE})) and its BO limit (i.e., eq.(\ref{eq:DE_BO})):

 \begin{eqnarray}\label{eq:DE2_F}
\nonumber \Delta E^{(2)}_{I0,OP}&=&\Delta E_{I0}^{f} +\\
\nonumber &+& \frac{1}{2} \sum_{ij} \left( \langle\Phi_I^f | \hat{V}_i |\Phi_I^f\rangle-\langle \Phi_0^f | \hat{V}_i| \Phi_0^f \rangle \right) \left( \langle\Phi_I^f | Q_{d,ij}\hat{V}_j^f |\Phi_I\rangle-\langle \Phi_0 | Q_{d,ij}\hat{V}_j| \Phi_0^f \rangle \right)\\
\nonumber &+& \frac{1}{2} \sum_{K\neq I} \sum_{ij}Q_{ij}'(\omega_{IK}) \langle\Phi_I^f | \hat{V}_i |\Phi_K^f\rangle\langle \Phi_K^f | \hat{V}_j| \Phi_I^f \rangle+\\
&-& \frac{1}{2} \sum_{K\neq 0} \sum_{ij}Q_{ij}'(\omega_{0K}) \langle\Phi_0^f | \hat{V}_i |\Phi_K^f\rangle\langle \Phi_K^f | \hat{V}_j| \Phi_0^f \rangle
\end{eqnarray}
 
\begin{eqnarray}\label{eq:DE_BO_F}
\Delta E_{I0,BO}^{(2)}&=&\Delta E_{I0} + \frac{1}{2} \sum_{ij} \langle\Phi_I^f| \hat{V}_i Q_{d,ij}  \hat{V}_j| \Phi_I^f \rangle -
\frac{1}{2} \sum_{ij} \langle\Phi_0^f| \hat{V}_i Q_{d,ij}  \hat{V}_j| \Phi_0^f \rangle
\label{eq:DE_BO_pol}
\end{eqnarray}
These equations have the same form as those for the apolar solvent, but calculated with frozen solvent states.

\section{ Derivation of the PCM expression for $E_{dis}$}
\label{sec:appA}

We provide here a derivation of eq.(\ref{eq:dis_int}) in the main text.
We re-write eq.(\ref{eq:dis}) as:
 \begin{eqnarray}
E_{dis}=\frac{1}{2}\sum_{K\neq 0,P\neq 0} \int d\vec{r}_1  \sum_{ij}\frac{1}{|\vec{r}_{1}-\vec{s}_i|} \rho(K0|\vec{r}_1) V_j(K0)\frac{\langle \Upsilon_0| \hat{q}_{j}|  \Upsilon_P \rangle \langle  \Upsilon_P | \hat{q}_{i}| \Upsilon_0 \rangle  }{\Delta E_{K0} + \Delta E_{P0}}
\label{eq:dis1}
\end{eqnarray}
where we have used that $V_j(K0)=\langle  \Phi_K | \hat{V}_{j}| \Phi_0 \rangle$ by definition and that:
\begin{eqnarray}
\langle  \Phi_0 | \hat{V}_{i}| \Phi_K \rangle = \int d\vec{r}_1  \frac{1}{|\vec{r}_{1}-\vec{s}_i|} \rho(K0|\vec{r}_1)
\end{eqnarray}
which is again related to the definition of the operator $\hat{V}_{i}$ and of the transition density $\rho(K0|\vec{r}_1)$. We now use  the following integral equation:\cite{Mcweeny1992methods}
\begin{eqnarray}
\frac{1}{\Delta E_{K0} + \Delta E_{P0}}=\frac{2}{\pi}\int_0^{\infty}d\omega ~\frac{\Delta E_{K0}}{\Delta E_{K0}^2+\omega^2}\frac{\Delta E_{P0}}{\Delta E_{P0}^2+\omega^2}
\label{eq:integ}
\end{eqnarray}
which holds for $\Delta E_{K0}>0$ and $\Delta E_{P0}>0$ to get:
 \begin{eqnarray}
\nonumber E_{dis}=\frac{1}{\pi}\int_0^{\infty}d\omega \sum_{K\neq 0} \frac{\Delta E_{K0}}{\Delta E_{K0}^2+\omega^2}  \int d\vec{r}_1  \sum_{ij}\frac{1}{|\vec{r}_{1}-\vec{s}_i|} \rho(K0|\vec{r}_1) V_j(K0)\\
\sum_{P\neq 0}\frac{\langle \Upsilon_0| \hat{q}_{j}|  \Upsilon_P \rangle \langle  \Upsilon_P | \hat{q}_{i}| \Upsilon_0 \rangle  \Delta E_{P0}}{\Delta E_{P0}^2+\omega^2}
\label{eq:dis2}
\end{eqnarray}
The last fraction of this expression can be further simplify by noting that:
 \begin{eqnarray}
\nonumber \sum_{P\neq 0}\frac{\langle \Upsilon_0| \hat{q}_{j}|  \Upsilon_P \rangle \langle  \Upsilon_P | \hat{q}_{i}| \Upsilon_0 \rangle  \Delta E_{P0}}{\Delta E_{P0}^2+\omega^2} = & & \\ \frac{1}{2}\sum_{P\neq 0} \frac{\langle \Upsilon_0| \hat{q}_{j}|  \Upsilon_P \rangle \langle  \Upsilon_P | \hat{q}_{i}| \Upsilon_0 \rangle }{\Delta E_{P0}+i\omega}+\frac{\langle \Upsilon_0| \hat{q}_{j}|  \Upsilon_P \rangle \langle  \Upsilon_P | \hat{q}_{i}| \Upsilon_0 \rangle }{\Delta E_{P0}-i\omega}=Q_{ij}(i\omega)
\label{eq:qq}
\end{eqnarray}
That leads to eq.(\ref{eq:dis_int}) in the main text. 

\section{ Derivation of real-frequency expression for $E_{dis}$}
\label{sec:appB}
The alternative expression for $E_{dis}$ given in eq.(\ref{eq:dis_cohen_2}) of the main text can be obtained by making use of the following integral relation:\cite{Cohen04}
\begin{eqnarray}
\nonumber \frac{1}{\Delta E_{K0} + \Delta E_{P0}}&=&\frac{1}{4}\int_{-\infty}^{\infty} d\omega \left( \left( \principalvalue \frac{1}{\Delta E_{K0} + \omega} + \principalvalue \frac{1}{\Delta E_{K0} - \omega} \right) \left( \delta (\omega+\Delta E_{P0}) + \delta (\omega-\Delta E_{P0}) \right) \right) +\\
&+&\left( \left( \principalvalue \frac{1}{\Delta E_{P0} + \omega} + \principalvalue \frac{1}{\Delta E_{P0} - \omega} \right) \left( \delta (\omega+\Delta E_{K0}) + \delta (\omega-\Delta E_{K0}) \right) \right)
\label{eq:int_cohen}
\end{eqnarray}
Using such expression, it is possible to reformulate $E_{dis}$ in eq.(\ref{eq:dis}) as: 
\begin{eqnarray}
\nonumber &E_{dis}&=\frac{1}{2}\sum_{K\neq0}\sum_{ij} \langle \Phi_0 | \hat{V}_{i}| \Phi_K \rangle \langle \Phi_K  | \hat{V}_{j}| \Phi_0 \rangle  Q'_{ij}(\Delta E_{K0})+\\
&+& \frac{1}{2}\int \frac{d\omega}{2\pi} \sum_{ij}\left( \sum_{K\neq 0} \principalvalue \frac{\langle 0| \hat{V}_{i}| \Phi_K  \rangle \langle \Phi_K | \hat{V}_{j} | 0 \rangle  }{\Delta E_{K0} + \omega} + \principalvalue \frac{\langle \Phi_0| \hat{V}_{i}| \Phi_K  \rangle \langle \Phi_K | \hat{V}_{j} | \Phi_0 \rangle  }{\Delta E_{K0} - \omega}\right) C_{ij}^{sym} (\omega) 
\label{eq:dis_cohen}
\end{eqnarray}
Eq.(\ref{eq:dis_cohen}) can be compacted to eq.(\ref{eq:dis_cohen_2}) in the main text by introducing the symmetric solute potential-potential correlation function and the solute potential response function and by noting that:
 \begin{eqnarray}
\frac{1}{2}\sum_{P\neq 0} \frac{\langle \Upsilon_0| \hat{q}_{j}|  \Upsilon_P \rangle \langle  \Upsilon_P | \hat{q}_{i}| \Upsilon_0 \rangle }{\Delta E_{P0}+\Delta E_{K0}}+  \frac{\langle \Upsilon_0| \hat{q}_{j}|  \Upsilon_P \rangle \langle  \Upsilon_P | \hat{q}_{i}| \Upsilon_0 \rangle }{\Delta E_{P0}-\Delta E_{K0}}=Q'_{ij}(\Delta E_{K0})
\label{eq:cohen1}
\end{eqnarray}
and, by definition  
\begin{eqnarray}
\nonumber \frac{\pi}{2} \sum_{P}\left( \langle \Upsilon_0| \hat{q}_{j}|  \Upsilon_P \rangle \langle  \Upsilon_P | \hat{q}_{i}| \Upsilon_0 \rangle + \langle \Upsilon_0| \hat{q}_{i}|  \Upsilon_P \rangle \langle  \Upsilon_P | \hat{q}_{j}| \Upsilon_0 \rangle \right) \left( \delta (\omega+\Delta E_{P0}) + \delta (\omega-\Delta E_{P0}) \right) &=&\\
=C_{ij}^{sym} (\omega) 
\label{eq:cohen2}
\end{eqnarray}
Note that in the last relation the term with $P=0$ is also included, which is however null as the expectation value of the charges in the solvent unperturbed ground state ($\langle \Upsilon_0| \hat{q}_{j}|  \Upsilon_0 \rangle$) is null (the solvent is not polarized when unperturbed).

\begin{acknowledgments}
S.C. would like to thank Silvio Pipolo and Emanuele Coccia for useful discussions. Funding from the EU H2020 ERC under the grant ERC-CoG-681285 TAME-Plasmons and by MIUR under the Grant R164LZWZ4A MIUR-FARE Plasmo-Chem is gratefully acknowledged. Computational work has been partially carried out on the C3P (Computational Chemistry Community in Padua) HPC facility of the Department of Chemical Sciences of the University of Padua.
\end{acknowledgments}

\bibliography{aipsamp}

\end{document}